\documentclass[preprint2]{aastex}

\slugcomment{Resubmitted to Astrophysical Journal in August 2014}

\shorttitle{Surface Geometry of Protoplanetary Disks}
\shortauthors{Takami et al.}

\begin{document}



\title{Surface Geometry of Protoplanetary Disks Inferred From Near-Infrared Imaging Polarimetry\footnote{Based on data collected at Subaru Telescope, which is operated by the National Astronomical Observatory of Japan.}}


\author{
Michihiro Takami\altaffilmark{1},
Yasuhiro Hasegawa\altaffilmark{1},
Takayuki Muto\altaffilmark{2}, 
Pin-Gao Gu\altaffilmark{1},
Ruobing Dong\altaffilmark{3,4,5}, 
Jennifer L. Karr\altaffilmark{1},
Jun Hashimoto\altaffilmark{6}, 
Nobuyuki Kusakabe\altaffilmark{7}, 
Edwige Chapillon\altaffilmark{1},
Ya-Wen Tang\altaffilmark{1},
Youchi Itoh\altaffilmark{8},
Joseph Carson\altaffilmark{9}, 
Katherine B. Follette\altaffilmark{10}, 
Satoshi Mayama\altaffilmark{11}, 
Michael Sitko\altaffilmark{12}, 
Markus Janson\altaffilmark{13}, 
Carol A. Grady\altaffilmark{14,15}, 
Tomoyuki Kudo\altaffilmark{16}, 
Eiji Akiyama\altaffilmark{7}, 
Jungmi Kwon\altaffilmark{7}, 
Yasuhiro Takahashi\altaffilmark{17,7},  
Takuya Suenaga\altaffilmark{18}, 
Lyu Abe\altaffilmark{19}, 
Wolfgang Brandner\altaffilmark{20}, 
Timothy D. Brandt\altaffilmark{21},  
Thayne Currie\altaffilmark{22}, 
Sebastian E. Egner\altaffilmark{16}, 
Markus Feldt\altaffilmark{20}, 
Olivier Guyon\altaffilmark{16}, 
Yutaka Hayano\altaffilmark{16}, 
Masahiko Hayashi\altaffilmark{7}, 
Saeko Hayashi\altaffilmark{16}, 
Thomas Henning\altaffilmark{20}, 
Klaus W. Hodapp\altaffilmark{23}, 
Mitsuhiko Honda\altaffilmark{24}, 
Miki Ishii\altaffilmark{7},  
Masanori Iye\altaffilmark{7},  
Ryo Kandori\altaffilmark{7}, 
Gillian R. Knapp\altaffilmark{21}, 
Masayuki Kuzuhara\altaffilmark{7,25,26}, 
Michael W. McElwain\altaffilmark{15}, 
Taro Matsuo\altaffilmark{27},
Shoken Miyama\altaffilmark{7},  
Jun-Ichi Morino\altaffilmark{7},  
Amaya Moro-Martin\altaffilmark{28}, 
Tetsuo Nishimura\altaffilmark{16}, 
Tae-Soo Pyo\altaffilmark{16},  
Eugene Serabyn\altaffilmark{29}, 
Hiroshi Suto\altaffilmark{7},  
Ryuji Suzuki\altaffilmark{7},  
Naruhisa Takato\altaffilmark{16},  
Hiroshi Terada\altaffilmark{16}, 
Christian Thalmann\altaffilmark{30}, 
Daigo Tomono\altaffilmark{16}, 
Edwin L. Turner\altaffilmark{21,31}, 
John P. Wisniewski\altaffilmark{6},  
Makoto Watanabe\altaffilmark{32}, 
Toru Yamada\altaffilmark{33}, 
Hideki Takami\altaffilmark{7},  
Tomonori Usuda\altaffilmark{7},  
Motohide Tamura\altaffilmark{7,17,18}  
}
\altaffiltext{1}{Institute of Astronomy and Astrophysics, Academia Sinica,
P.O. Box 23-141, Taipei 10617, Taiwan, R.O.C.; hiro@asiaa.sinica.edu.tw}
\altaffiltext{2}{Division of Liberal Arts, Kogakuin University, 1-24-2, Nishi-Shinjuku, Shinjuku-ku, Tokyo, 163-8677, Japan} 
\altaffiltext{3}{Nuclear Science Division, Lawrence Berkeley National Laboratory, 1 Cyclotron Road, Berkeley, CA, 94720, USA} 
\altaffiltext{4}{Departments of Astronomy, University of California, Berkeley, CA 94720, USA} 
\altaffiltext{5}{Hubble Fellow} 
\altaffiltext{6}{H.L. Dodge Department of Physics and Astronomy, University of Oklahoma, 440 W Brooks St Norman, OK 73019, USA} 
\altaffiltext{7}{National Astronomical Observatory of Japan, 2-21-1 Osawa, Mitaka, Tokyo 181-8588, Japan} 
\altaffiltext{8}{Nishi-Harima Astronomical Observatory, Center for Astronomy, University of Hyogo, 407-2 Nishigaichi, Sayo, Sayo, Hyogo 679-5313, Japan} 
\altaffiltext{9}{Department of Physics and Astronomy, College of Charleston, 58 Coming St., Charleston, SC 29424, USA} 
\altaffiltext{10}{Steward Observatory, University of Arizona, 933 N Cherry Ave, Tucson AZ 85721} 
\altaffiltext{11}{The Center for the Promotion of Integrated Sciences, The Graduate University for Advanced Studies~(SOKENDAI), Shonan International Village, Hayama-cho, Miura-gun, Kanagawa 240-0193, Japan} 
\altaffiltext{12}{Department of Physics, University of Cincinnati, Cincinnati OH 45221, USA} 
\altaffiltext{13}{Astrophysics Research Center, Queen's University Belfast, BT7 1NN, Northern Ireland, UK} 
\altaffiltext{14}{Eureka Scientific, 2452 Delmer Suite 100, Oakland CA 96402, USA} 
\altaffiltext{15}{ExoPlanets and Stellar Astrophysics Laboratory, Code 667, Goddard Space Flight Center, Greenbelt, MD 20771, USA} 
\altaffiltext{16}{Subaru Telescope, 650 North AÕohoku Place, Hilo, HI 96720, USA}
\altaffiltext{17}{Department of Astronomy, The University of Tokyo, 7-3-1 Hongo, Bunkyo-ku, Tokyo 113-0033, Japan} 
\altaffiltext{18}{Department of Astronomical Science, The Graduate University for Advanced Studies (SOKENDAI), 2-21-1 Osawa, Mitaka, Tokyo 181-8588, Japan} 
\altaffiltext{19}{Laboratoire Lagrange (UMR 7293), Universit\`{e} de Nice-Sophia Antipolis, CNRS,
Observatoire de la C\^{o}te d'Azur, 28 avenue Valrose, 06108 Nice Cedex 2, France} 
\altaffiltext{20}{Max Planck Institute for Astronomy, Koenigstuhl 17, D-69117 Heidelberg, Germany}
\altaffiltext{21}{Department of Astrophysical Sciences, Princeton University, Peyton Hall, Ivy Lane, Princeton, NJ 08544, USA}
\altaffiltext{22}{Department of Astronomy and Astrophysics, University of Toronto, Toronto, ON, Canada} 
\altaffiltext{23}{Institute for Astronomy, University of Hawaii, 640 North AÕohoku Place, Hilo, HI 96720, USA} 
\altaffiltext{24}{Department of Mathematics and Physics, Faculty of Science, Kanagawa University, 2946 Tsuchiya, Hiratsuka, Kanagawa 259-1293, Japan} 
\altaffiltext{25}{Department of Earth and Planetary Science, The University of Tokyo, 7-3-1 Hongo, Bunkyo-ku, Tokyo 113-0033, Japan} 
\altaffiltext{26}{Department of Earth and Planetary Sciences, Tokyo Institute of Technology, 2-12-1 Ookayama, Meguro-ku, Tokyo 152-8551, Japan} 
\altaffiltext{27}{Department of Astronomy, Kyoto University, Kitashirakawa-Oiwake-cho, Sakyo-ku, Kyoto, Kyoto 606-8502, Japan} 
\altaffiltext{28}{Department of Astrophysics, CAB-CSIC/INTA, 28850 Torrej—n de Ardoz, Madrid, Spain} 
\altaffiltext{29}{Jet Propulsion Laboratory, California Institute of Technology, Pasadena, CA, 91109, USA} 
\altaffiltext{30}{Institute for Astronomy, ETH Zurich, Wolfgang-Pauli-Strasse 27, 8093, Zurich, Switzerland} 
\altaffiltext{31}{Kavli Institute for the Physics and Mathematics of the Universe, The University of Tokyo, Kashiwa 277-8568, Japan} 
\altaffiltext{32}{Department of Cosmosciences, Hokkaido University, Kita-ku, Sapporo, Hokkaido 060-0810, Japan} 
\altaffiltext{33}{Astronomical Institute, Tohoku University, Aoba-ku, Sendai, Miyagi 980-8578, Japan} 

\begin{abstract}
We present a new method of analysis for determining the surface geometry of five protoplanetary disks observed with near-infrared imaging polarimetry using Subaru-HiCIAO. Using as inputs the observed distribution of polarized intensity ($PI$), disk inclination, assumed properties for dust scattering, and other reasonable approximations, we calculate a differential equation to derive the surface geometry. This equation is numerically integrated along the distance from the star at a given position angle. We show that, using these approximations, the local maxima in the $PI$ distribution of spiral arms (SAO 206462, MWC 758) and rings (2MASS J16042165--2130284, PDS 70) are associated with local concave-up structures on the disk surface. We also show that the observed presence of an inner gap in scattered light still allows the possibility of a disk  surface that is parallel to the light path from the star, or a disk that is shadowed by structures in the inner radii.
Our analysis for rings does not show the presence of a vertical inner wall as often assumed in studies of disks with an inner gap.  
Finally, we summarize the implications of spiral and ring structures as potential signatures of ongoing planet formation.
\end{abstract}


\keywords{protoplanetary disks --- stars: individual (SAO 206462, MWC 758, 2MASS J16042165--2130284, PDS 70, MWC 480) --- stars: pre-main sequence --- polarization}



\section{INTRODUCTION}

Structures in protoplanetary disks are of particular interest for identifying and investigating planet formation. Coronagraphic imaging at optical and near-infrared wavelengths has been extensively used to reveal structures in disks at the highest angular resolutions currently available ($\sim$ 0\farcs1). In particular, observations of scattered light, enabled through the technique of coronagraphic imaging, can be used to trace the presence ofÊ dust grains in the vicinity of the central star.

Recently, a large survey program, ``Strategic Explorations of Exoplanets and Disks with Subaru" \citep[SEEDS,][]{Tamura09}, has been conducted \citep[e.g.,][]{Thalmann10,Kusakabe12,Tanii12,Follette13} with the dedicated coronagraphic imager Subaru-HiCIAO \citep{Tamura06_HiCIAO} and AO188 \citep{Hayano04}. Polarized intensity ($PI$) images have been taken for most of the objects in the SEEDS disk survey program, as such \textit{PI} images suffer significantly less contamination from the stellar flux, compared with Stokes $I$ images. The SEEDS survey has revealed spiral structures, inner gaps, and  azimuthal gaps in ring-like flux distributions in some disks
\citep[e.g.,][]{Hashimoto11,Muto12,Mayama12}.
These investigations interpret the observed structures as potential signatures of ongoing planet formation.  The investigations from the SEEDS survey have been paralleled by efforts on other telescopes, sometimes employing similar near-infrared $PI$ imaging
\citep[e.g.,][]{Avenhaus14}.

Some of the SEEDS studies have assumed that the observed spiral and ring structures approximately match the surface density distribution \citep[][]{Muto12,Hashimoto12,Mayama12,Grady13}. However, some discrepancies in flux distributions have been reported  between the SEEDS observations and the millimeter/sub-millimeter dust continuum \citep{Mayama12,Tang12,Muto12,Grady13}. Young disks are optically thick in the near-infrared \citep[e.g.,][for a review]{Watson07_PPV} but thin in the millimeter/sub-millimeter continuum \citep[e.g.,][for reviews]{Williams11,Espaillat14}; therefore, the latter more directly traces the surface density distribution. This implies that we may have to revise our interpretations of spiral structures, inner gaps, and/or azimuthal gaps seen in the near-infrared.

Observed scattered light from disks has also been compared with monochromatic and full radiative transfer simulations \citep[e.g.,][]{Duchene04,Dong12b,Dong12a,Grady13,Follette13,Takami13}.
Scattering occurs at the surface of the protoplanetary disks; hence, the observed scattered light is not a direct or unique function of the density distribution parameters used in these simulations. For instance, \citet{Takami13} show that the modeled $PI$ distributions of the disk surface are relatively independent of the surface density distribution and scale height once we define a disk radius and an optical thickness in a given direction from the star. Considering a gravitationally collapsing disk and a disk with a gap,
\citet{Jang-Condell07},
\citet{Jang-Condell12,Jang-Condell13}, and
\citet{Muto11}
discuss a complex relationship between the disk surface and the parameters for internal density distribution. 

Employing a few reasonable approximations, we have developed an alternative simplified method to visualize disk surface geometries. We apply it to SEEDS observations that have exhibited either spiral structures \citep[SAO 206462, MWC 758 ---][]{Muto12,Grady13}, a ring \citep[2MASS J16042165--2130284, PDS 70  ---][]{Mayama12,Hashimoto12,Dong12b}, or a relatively uniform $PI$ distribution \citep[MWC 480 ---][]{Kusakabe12}.
In Section 2 of this paper we present a differential equation to derive the surface geometry of the disk from the observed $PI$ distribution of scattered light.
In Section 3 we perform radiative transfer simulations to verify our new approach. With simulated $PI$ images that are generated for more realistic disk models, we confirm that the surface geometry computed by our technique corresponds very well to the one defined by the density distribution of the models.
In Section 4 we describe the surface geometry obtained for the aforementioned disks.
In Section 5 we discuss possible implications of our results for spiral and ring structures, and also summarize possible mechanisms for non-axisymmetric illumination in some disks.
In Section 6 we give conclusions.


\section{FORMULA TO DERIVE THE SURFACE GEOMETRY}

\begin{figure*}
\epsscale{1.8}
\plotone{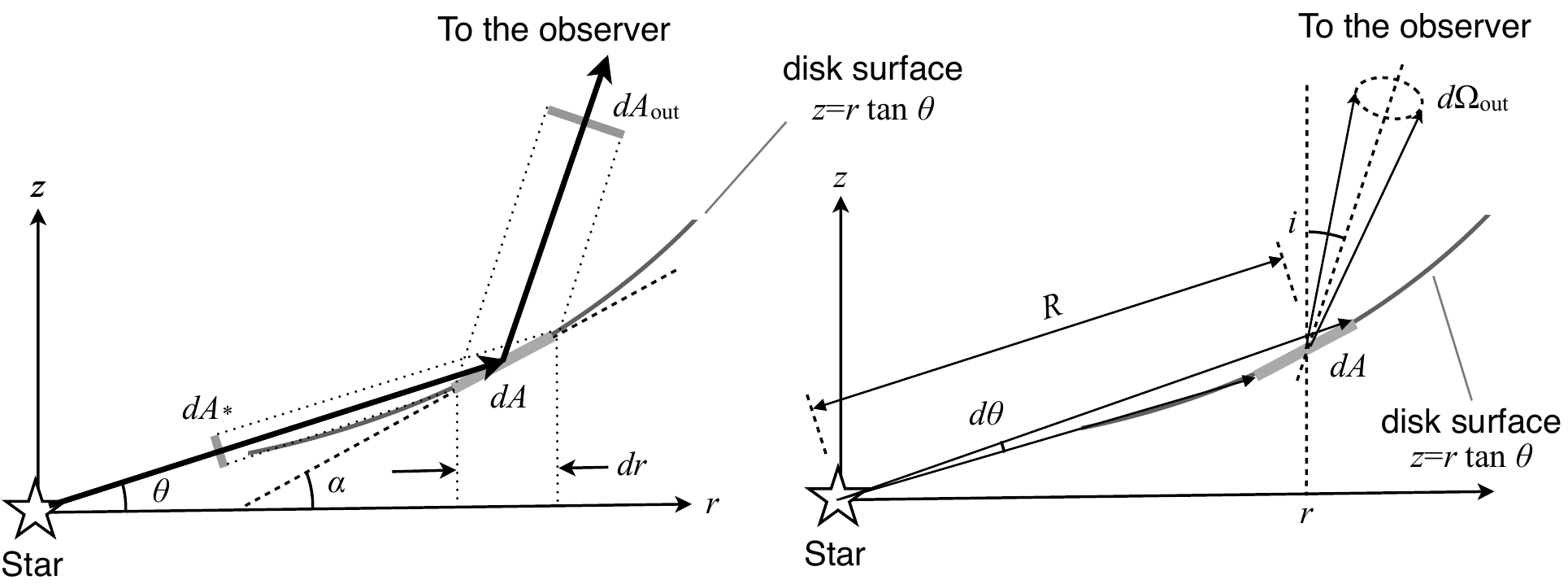}
\caption{Geometry and labels for the disk.
See Table \ref{tbl_params} for details of individual parameters.
\label{fig_geometry}}
\end{figure*}

Figure \ref{fig_geometry} shows a diagram of the scattering geometry. In Table \ref{tbl_params} we summarize parameters used in this section.

In Section 2.1 we describe approximations and assumptions used to derive the formula.
In Section 2.2 we derive the formula to determine the surface geometry. In Section 2.3 we describe the numerical integration of the formula and discuss two of the approximations and assumptions.

\subsection{Approximations and Assumptions}

\begin{enumerate}
\item The analysis below applies to disks for which the surface emission can be observed, with the viewing the angle $i$ not close to 90$^\circ$ (the edge-on view).

\item The disk is optically thick in the near-infrared, at least for the wavelength range encompassed by our observational data \citep[see][for a review]{Watson07_PPV}. This fact has been observationally confirmed in some young edge-on disks, where the disk appears as a dark lane \cite[e.g.,][]{Padgett99}, and in disks silhouetted against bright nebular backgrounds \citep[see][ for a review]{McCaughrean00}.
The optically thick nature of the disk is also consistent with some simulations that model near-infrared scattered light from protoplanetary disks \citep[e.g.,][]{Silber00,Duchene04,Dong12b}.

\item A disk surface is defined by the locus of points, between the star and outer disk radius, with an optical depth of approximate unity  \citep{Watson07_PPV,Jang-Condell07,Jang-Condell08,Perrin09,Muto11}. The width of the scattering layer on the disk is significantly smaller than the geometrical thickness of the disk. This approximation has been used for theoretical determinations of the radiative heating of the disk by the central star \citep[][]{Kenyon87,Chiang97}. \citet{Takami13} numerically examined this issue for some disk models, finding that the positions of the $\tau$=0.5, 1 and 2 curves approximately match, agreeing with the above approximation. Observationally, this approximation is corroborated by optical and near-infrared observations of some edge-on disks, which show a clear boundary between the shadowed flared disk and the reflection nebulae \citep[e.g.,][]{Padgett99,Watson07_PPV}. 

We note that the above definition of the disk surface does not allow the exploration of regions of the disk which do not receive any photons, or in other words, regions of the disk where we cannot define a position corresponding to an optical thickness $\tau \sim 1$. These cases include (a) a disk surface parallel to the light path from the star; (b) areas outside the disk where dust does not exist, or regions shadowed by the outer edge of the disk; or (c) regions shadowed by an inner disk structure (e.g., a puffed-up inner rim due to heating from stellar radiation \citep[e.g.,][]{Dullemond01}
or by the protruding inner edge of a disk gap \citep[e.g.,][see also Section 3]{Jang-Condell12,Jang-Condell13}.

We also note that we cannot accurately determine the disk surface if an optically thin layer above the disk (e.g., a remnant envelope) contribute to the optical thickness between the star and the disk surface \citep[e.g.,][]{Stapelfeldt03,Follette13,Takami13}.

\item The disk is geometrically thin ($z/r \ll 1$, where $z$ and $r$ are the height from the midplane and the distance from the star in the midplane, respectively). This assumption is reasonable for most protoplanetary disks \citep[see][for reviews]{Watson07_PPV,Dullemond07_PPV}. We note that a ``geometrically thin disk'' does not always imply that the model disk is flat \citep[e.g.,][]{Whitney92}. For instance, a variation in $z$ or $z/r$, including a concave-up or concave-down geometry, is still possible with the constraint $z/r \ll 1$ (see Sections 3 and 4 for examples).

\item As shown in Figure \ref{fig_geometry}, $\alpha$ is the angle of the surface of the disk at the position of scattering with respect to the midplane. $\theta$ is the elevation angle of this position from the midplane. We assume that $\alpha$ is not significantly greater than $\theta$ (i.e., $\alpha - \theta \ll 1$). This is not a standard assumption for protoplanetary disks but we confirm its validity for our sources in Sections 3 and 4. We note that $\alpha$ is always equal to or larger than $\theta$ due to the definition of the surface described above.

\item The star is treated as a point source of illumination. In practice, a pre-main sequence star has a stellar radius of a few solar radii \citep{Stahler05}. As shown below, this approximation does not cause significant errors.

\item The contribution of photons, with multiple scattering, to the $PI$ flux is negligible. This approximation is supported by \citet{Takami13}.  For the scattering geometries used in that study, the contribution of multiply scattered photons to the $PI$ flux was within 10 \% of that for singly scattered photons. This is because, in the case of small grains, photons are relatively isotropically scattered and  the scattered photons have polarizations with a variety of position angles (P.A.s), thus canceling each other out. In the case of large grains, most of the photons are scattered forward, resulting in a significant reduction in individual photon polarization after the first scattering.

\end{enumerate}

\subsection{Derivation}

In the case where the contribution of multiple scatterings is negligible (Approximation 7), the scattered $PI$ intensity from the disk surface is described as follows:
\begin{equation}
I_{\nu; \rm out} \it (r,\phi) ~dA_{\rm out} ~d \rm\Omega_{out} \it = \frac{L_\nu}{\rm 4 \pi \it R^{\rm2} \it (r,\phi)} \left( \frac{\it PI}{I_{\rm 0}} \right) dA_* d \rm\Omega_{out},
\end{equation}
where $I_{\nu; \rm out}\it (r,\phi)$ is the observed $PI$ intensity at the distance $r$ from the star and the azimuthal angle $\phi$ in the plane of the disk, $dA_{\rm out}$ is the differential area of the disk surface ($dA$) projected onto the plane of the sky, $d \rm\Omega_{out}$ is the solid angle from the disk surface toward the observer, 
and  $L_\nu$ is the stellar luminosity per unit frequency. $R(r,\phi)$ is the distance from the star to the surface of the disk, i.e., the position of a scattering event. This parameter is equal to $r$/cos $\theta(r,\phi)$, where $\theta (r,\phi)$ is the elevation angle of the position of a scattering event from the midplane. ($PI/I_0$) is the fraction of the scattered $PI$ intensity relative to the incident $I_0$ flux (sr$^{-1}$). $dA_*$ is the differential area $dA$ projected on the plane perpendicular to the light path from the star to the disk.

The right side of Equation (1) is a product of the incident light on a unit differential area at the disk surface ($L_\nu dA_* / 4 \pi R^2$) and the fraction of the energy radiated in the given direction (($PI/I_0)~d\Omega_{\rm out}$).
The parameter ($PI/I_0$)  in Equation (1) depends on the dust particles and the scattering angle \citep[Figure 5 of][]{Takami13}. For non-polarimetric disk imaging at optical and near-infrared wavelengths, this ratio would be replaced by the ``phase function''.  

The differential area of the disk surface projected onto the plane of the sky $dA_{\rm out}$ is:
\begin{equation}
dA_{\rm out} = dA \cos (i+\alpha),
\end{equation}
where $i$ is the disk inclination or the viewing angle.

The parameter $dA_*$, the differential area projected on the plane perpendicular to the light path from the star to the disk, is:
\begin{equation}
dA_* = dA \sin (\alpha - \theta).
\end{equation}

Substituting Equations (2) and (3) into Equation (1), we derive:
\begin{equation}
\left( \cos i - \sin i \tan \alpha \right) {I_{\nu; \rm out} }=\frac{L_\nu}{4\pi R^2} \left( \frac{\it PI}{I_{\rm 0}} \right) (\tan \alpha \cos \theta - \sin \theta). \label{eq:scat}
\end{equation}
Since $z/r=\tan \theta$ and $dz/dr=\tan \alpha$, where $z$ is the height of the disk surface from the midplane,
\begin{equation}
\tan\alpha=dz/dr = \tan \theta + (r/\cos^2 \theta) d\theta /dr.\label{eq:curv}
\end{equation}
Substituting Equation (5) into Equation (4) we derive:
 \begin{equation}
{d\theta \over dr}={\cos i - \sin i \tan \theta \over C \sin i + \cos \theta}{4\pi r I_{\nu,out} \over L_\nu (PI/I_0)},
\end{equation}
where 
\begin{equation}
C \equiv \frac{4\pi R^2 I_{\nu; \rm out}}{L_\nu (PI/I_0)} = \frac{dA_*}{dA_{\rm out}} = \frac{\sin (\alpha - \theta)}{\cos (i+\alpha)}
\end{equation}
As described above, $0 \leq \alpha - \theta \ll 1$ (Approximation/Assumption 5) and the disk is not close to edge-on (i.e., $i \gg \alpha$, Approximation/Assumption 1), therefore $C \ll 1$. Furthermore, 
the disk is geometrically thin ($\tan \theta \ll 1$, Approximation/Assumption 4) and the disk is not close to edge-on (i.e., not $\cos i \ll \sin i$, Approximation/Assumption 1), therefore $\cos i \gg \sin i \tan \theta$ in Equation (6). Therefore Equation (6) is approximated by:
%
\begin{eqnarray}
dS & \sim & \frac{ \rm 4 \pi \it r I_{\nu; \rm out }(r,\phi)  \rm ~cos \it ~i }{L_\nu (PI/I_0)} dr.
\end{eqnarray}
where $S$=sin $\theta$.

Using the following respective equations, we can replace $I_{\nu; \rm out} \it (r,\phi)$ and $L_\nu$ with the observed number of photons per second ($n_{\nu;PI}$, $n_{\nu;*}$):
\begin{eqnarray}
n_{\nu;PI} (r,\phi)& = &\frac{I_{\nu; \rm out} \it (r,\phi)}{h\nu} A_{\rm pix} \Omega_{\rm Tel} , \nonumber \\
n_{\nu;*}     & = & \frac{1}{h \nu} \it \frac{L_\nu}{\rm 4 \pi} \rm \Omega_{Tel}, \nonumber \\
\frac{n_{\nu;PI} (r,\phi)}{n_{\nu;*}} &=& \frac{\rm 4 \pi \it I_{\nu; \rm out} \it (r,\phi) A_{\rm pix}}{L_\nu}
\end{eqnarray}
where $h\nu$ is the photon energy;  $A_{\rm pix}$ is the area corresponding to the HiCIAO pixel scale (9.5 mas) at the distance to the target; $\Omega_{\rm Tel}$ is the solid angle corresponding to the telescope mirror ($\Omega_{\rm Tel} = A_{\rm Tel}/4 \it \pi d^{\rm 2}$, where $A_{\rm Tel}$ and $d$ are the area of the telescope mirror and the distance to the target, respectively). Substituting Equation (9) in Equation (8), we derive:
\begin{eqnarray}
dS & \sim & r \frac{n_{\nu;PI}(r,\phi)}{n_{\nu;*}} \frac{\rm ~cos \it ~i }{A_{pix}} \left( \frac{PI}{I_{\rm 0}} \right)^{-1} dr,
\end{eqnarray}

The surface geometry $z(r,\phi) = r$ tan [$\theta(r,\phi)$] can then be derived via numerical integration of $dS$, using Equation (10), along a given azimuthal angle in the disk coordinate ($\phi$) or the observed P.A.
The input parameters are the observed $PI$ distribution normalized to the stellar $I$ flux ($n_{PI}/n_{*}$), the inclination $i$, and an assumed dust property ($PI/I_0$).
For a given $\phi$ or P.A., the scattering angle should be approximately constant over any position if the disk is geometrically thin. Thus, $(PI/I_0)$ is approximately constant if the dust properties do not vary along with P.A. For simplicity, we show all the results in later sections in terms of P.A.s of the observations, rather than azimuthal angles $\phi$ in the disk coordinate.

\subsection{Numerical Integration}

We use the extended Newton-Cotes formula \citep{Press92} to numerically integrate Equation (10).
We tested this algorithm as follows. First, we provided the surface geometry using an analytic function and analytically derived the $PI$ distribution. Then we performed the numerical integration and investigated if the calculated surface geometry matches the original geometry.
We found that the surface geometry is reproduced well.

The initial value of the numerical integration $S_0$ (=sin $\theta_0$, where $\theta_0$ is the elevation angle of the position of a scattering event from the midplane) can be arbitrarily given at radius $r_0$.
This convention corresponds to an arbitrary inclination of the disk surface with respect to the midplane.
The observed flux is independent of $S_0$ or $\theta_0$ because an arbitrary inclination will not change the variation in the illumination of the disk.


As described above, we have used the assumption of $0 \leq \alpha - \theta \ll 1$ (where $\alpha$ is the angle of the surface with respect to the midplane and $\theta$ is the elevation angle of the position of a scattering event from the midplane), which may not be applicable for all cases. This can be verified by investigating whether $C \ll 1$ using Equation (7). Substituting Equations (9) to Equation (7) we derive:
\begin{eqnarray}
C = \frac{n_{\nu;PI}(r,\phi)}{n_{\nu;*}} \frac{R^2}{A_{\rm pix}} \left( \frac{PI}{I_{\rm 0}} \right)^{-1}, \nonumber
\end{eqnarray}
or
\begin{equation}
C \sim \frac{n_{\nu;PI}(r,\phi)}{n_{\nu;*}} \frac{r^2}{A_{\rm pix}} \left( \frac{PI}{I_{\rm 0}} \right)^{-1}.
\end{equation}
if the disk is geometrically thin (i.e., $r \sim R$). Therefore, we can determine whether $C \ll 1$ using the observed parameters ($n_{\nu;PI}/n_{\nu;*}$, $r$), the instrument parameter ($A_{\rm pix}$) and the scattering properties based on the dust model ($PI/I_0$).

To derive Equation (10), we regard the star as the point source of illumination. In practice, a pre-main sequence star has a stellar radius of a few solar radii \citep{Stahler05}. Given the considered angular radius of the star from the disk surface, the point-source approximation yields an error on order of $10^{-4}$ for $\theta$, $S$, and $z/r$, at 100 AU. This error is significantly smaller than the disk structure scales we discuss in later sections.


\section{Comparison with Monochromatic Radiative Transfer Simulations}

We test the method of Section 2 using several disk models and our code for monochromatic Monte-Carlo simulations of scattered light. Our investigations are made as follows: (I) we define a density distribution and calculate the positions for $\tau$=0.5, 1, and 2 from the star (i.e., the location of the scattering layer) ; (II) we calculate the $PI$ distribution via simulation; and (III) we derive the surface geometry using (II) and Equation (10), and perform comparisons with (I). 

The modeled system consists of an illumination source with an axisymmetric circumstellar disk. The density distribution of a standard flared accretion disk \citep[e.g.,][]{Shakura73,Lynden-Bell74} is described in cylindrical coordinate ($r$,$z$) by:
\begin{equation}
\rho (r,z) = \rho_0 \left[1-\sqrt{\frac{R_*}{r}} \right] \left(\frac{R_*}{r} \right)^\alpha ~ \rm exp~ \left\{- \frac{1}{2} \left[\frac{\it z}{\it h}\rm \right]^2 \right\},
\end{equation}
where $\rho_0$ is a constant to scale the density, $R_*$ is the stellar radius, $\alpha$ is the radial density exponent, and $h$ is the disk scale height. The scale height $h$ increases with radius as $h = h_0 r^\beta$, where $\beta$ is the flaring index ($\beta > 0$). This model is commonly used by other researchers \citep[e.g.,][]{Cotera01,Whitney03a,Robitaille06,Follette13,Takami13}.
We make the common assumption $\alpha = \beta+1$ \citep[e.g.,][]{Robitaille06,Robitaille07,Dong12b,Dong12a,Follette13,Takami13}. This yields a surface density distribution $\Sigma (r) \propto r^{-1}$, which is generally consistent with that inferred from millimeter interferometry for disks associated with many low-mass pre-main sequence stars \citep{Andrews09,Andrews10b}. In addition to the parameters of Equation (12) the minimum and maximum radii of the disk are included as free parameters. 

\begin{figure*}
\epsscale{2.1}
\plotone{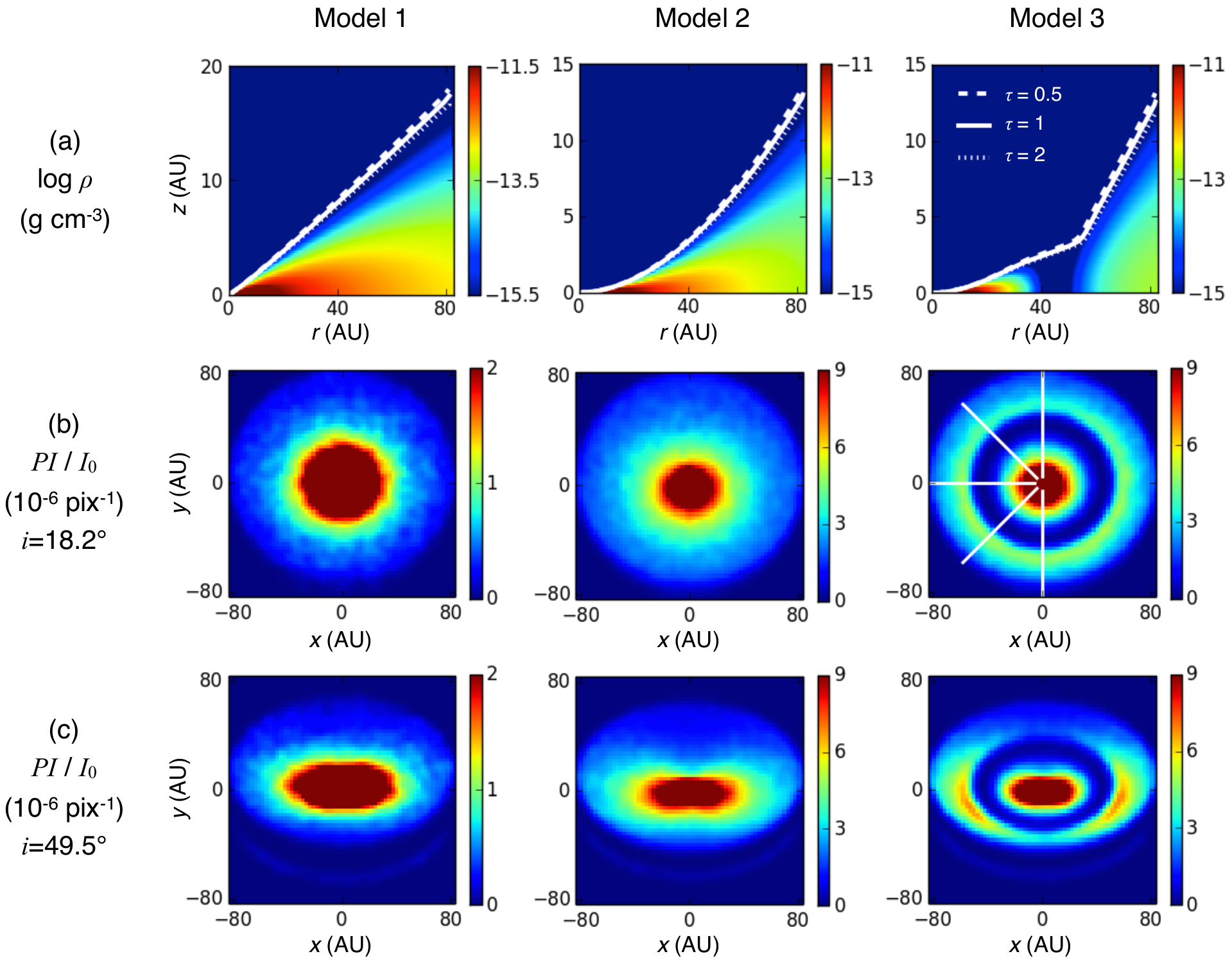}
\caption{
(a) Density distributions assumed for the ID 300949 model (Model 1), the same model but $\beta$=2, $\alpha$=3 (Model 2), and same as Model 2 but with a disk gap (Model 3, see text for details). White dashed, solid, and dotted curves show the positions for $\tau$=0.5, 1, and 2 from the star, respectively. (b)(c) Modeled $PI$ images using the Sprout code with a viewing angle of 18.2$^\circ$ and 49.5$^\circ$, respectively. White lines show the positions where we measured the $PI$ fluxes to calculate the surface geometry of the disk (P.A.=0, 45, 90, 135, and 180$^\circ$). 
\label{fig_vs_Sprout_images}}
\end{figure*}

Our calculations are made using the following models: (1) Model 300949 of \citet{Robitaille07}, which provides the best fit for the spectral energy distribution of RY Tau \citep{Takami13} (see Table \ref{tbl_3000949} for the list of parameters); (2) same as (1) but with large flaring ($\beta=2$); (3) same as (2) but with the addition of the term \{cos$^4$ [($r$/90 AU)$\pi$]\} to the right side of Equation (12) to create a disk gap (see Figure \ref{fig_vs_Sprout_images}). We do not include the envelope of the original model of \citet{Robitaille07}. \citet{Takami13} carried out simulations with and without the envelope, and found that the presence of the envelope does not affect the scattered flux.

For the dust opacity and scattering we use a standard interstellar dust model with homogeneous spherical particles \citep[see e.g.,][]{Cotera01,Wood02b,Robitaille06,Dong12a,Takami13}. 
The grain compositions are astronomical silicate and carbon dust without an ice coating, using graphite for the carbon dust\footnote{Different authors use different types of carbon dust, either graphite or amorphous carbon \citep{Cotera01,Wood02b}. While graphite has been extensively used \citep[e.g.,][]{Draine84,Laor93,Kim94,Whitney03a,Robitaille06,Dong12b,Dong12a}, far-infrared SEDs of young stellar objects and evolved stars suggest the absence of graphite and presence of amorphous carbon in circumstellar dust \citep[][and references therein]{Jager98}. We continue to use graphite for the KMH distribution for consistency with the authors as their size distribution is determined assuming graphite for the carbon dust.}. The size distribution of the dust grains is that of interstellar dust as measured by \citet{Kim94} ($R_V=3.1$, hereafter KMH).
The optical constants for astronomical silicate and graphite are obtained from \citet{Draine84}. The detailed scattering properties of this interstellar dust model are discussed in \citet[][]{Takami13}.
The gas-to-dust mass ratio is assumed to be 100.

The monochromatic Monte-Carlo scattering simulations were made using the Sprout code\footnote{Available at http://www.asiaa.sinica.edu.tw/\~{}jkarr/Sprout/sprout.html} \citep{Takami13}, which follows the method described in \citet{Fischer94}. We place an unpolarized point source at the centre the disk to the represent the star.
Multiple scatterings in the disk are included. The light path for the next scattering position is calculated for an opacity distribution based on the disk in Equation (12) and the dust opacity ($\kappa_{ext}=36$ cm$^{-2}$ g$^{-1}$ for the above dust model). The scattering angle and Stokes parameters after scattering are calculated using Mie theory. The Stokes parameters for each photon are initially set to ($I_0$,$Q_0$,$U_0$,$V_0$)=(1,0,0,0) and normalized to $\it I_{\rm out} \it = albedo \cdot \it I_{\rm in}$ after each scattering. The calculations for Mie scattering are made using the code developed by \citet{Wiscombe96}.

We use $10^7$ photons for each simulation. The photons escaping from the disk are collected in two imaginary detectors at viewing angles of 18.2$^\circ$ and 49.5$^\circ$. The disk is placed at a distance of 140 pc from the detector. Photons which do not interact with the disk are not collected. In order to normalize the $PI$ flux to the Stokes $I$ flux of the star, we separately calculate the expected number of photons for each viewing angle based on the incident number of photons and extinction. The pixel scale is set to twice that of Subaru-HiCIAO (9.5 mas) to increase the signial-to-noise. We divided the modeled $PI/I_*$ (where $I_*$ is the stellar $I$ flux) distribution by 4 to apply Equation (10)Ä with the pixel scale of HiCIAO. We then convolved the $PI/I_*$ images with a gaussian with a FWHM of 2.5 pixel, corresponding to $\sim$0\farcs05, to match the observations with Subaru-HiCIAO.

Figure \ref{fig_vs_Sprout_images} shows the density distributions for these disk models and the $PI$ images calculating using the Sprout code at viewing angle of 18.2$^\circ$ and 49.5$^\circ$. For Equation (10) the $PI$ distribution is extracted starting at the projected radius of 5.3 AU (2 pixels) from the star along P.A.=0$^\circ$, 45$^\circ$, 90$^\circ$, 135$^\circ$ and 180$^\circ$. Table \ref{tbl_PI_I0} shows a typical scattering angle corresponding to each viewing angle and P.A., calculated with a geometrically thin approximation. The table also shows ($PI/I_0$) (one of the constants in Equation 10) corresponding to these scattering angles based on the above dust model.

\begin{figure*}
\epsscale{1.8}
\plotone{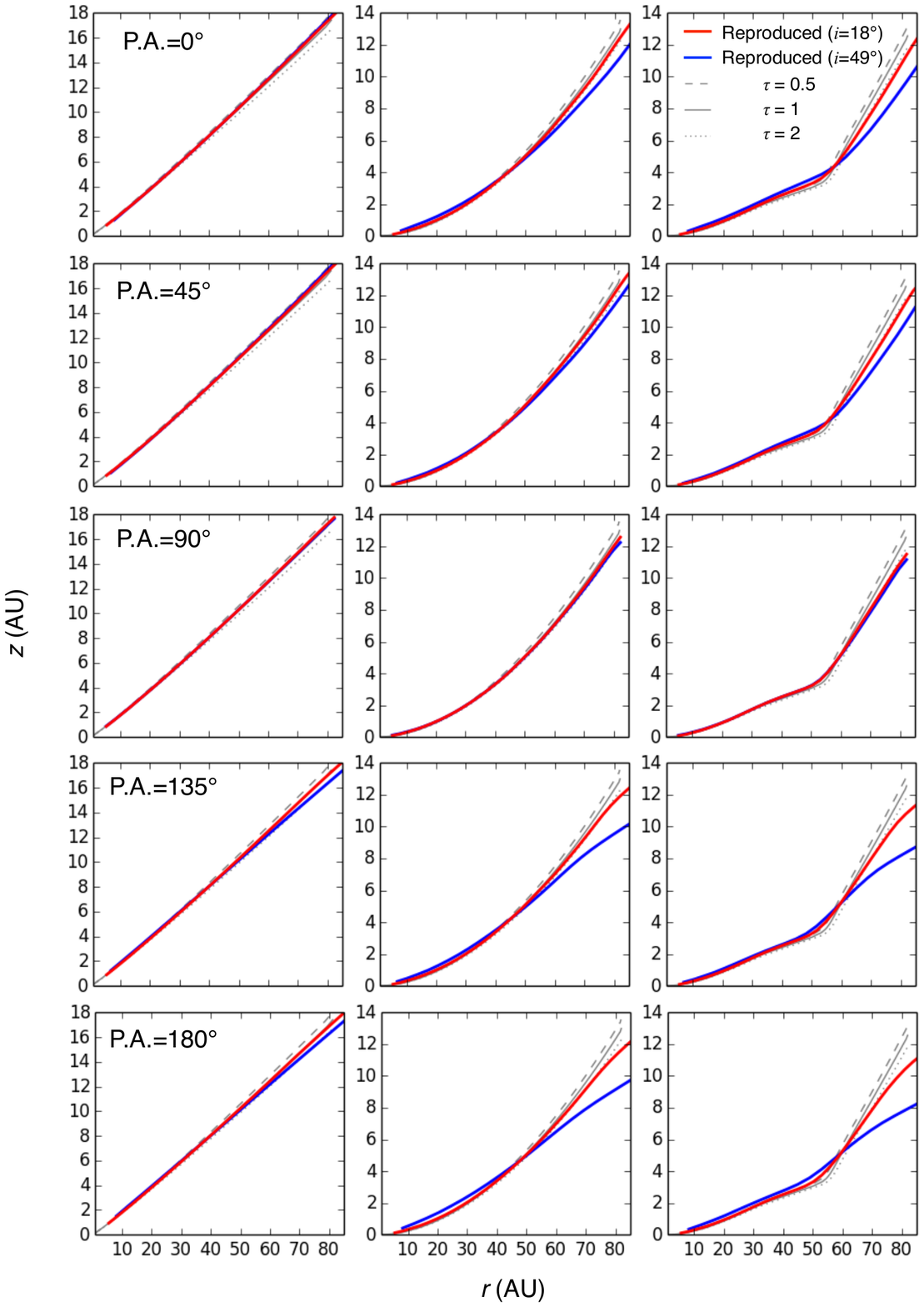}
\caption{
The surface geometry of the disk derived using the measured $PI$ flux in Figure \ref{fig_vs_Sprout_images} and Equation (10). The parameters $z$ and $r$ are the height from the midplane and the distance from the star in the midplane, respectively. The red and blue curves show the geometry derived from the $PI$ images with viewing angles of 18$^\circ$ and 49$^\circ$, respectively. Gray dashed, solid, and dotted curves show the positions for $\tau$=0.5, 1, and 2 from the star, respectively. The parameter $S_0$ in each case is chosen for the best fit of the $\tau$=1 curve (see text).
\label{fig_vs_Sprout_z_vs_r}}
\end{figure*}

Figure \ref{fig_vs_Sprout_z_vs_r} shows the surface geometries for the individual disks derived using Equation (10). These are plotted together with the positions for $\tau$=0.5, 1, and 2 from the star derived by numerical integration of the opacity distribution. The parameter $S_0$ is adjusted for the best fit of the results to the  $\tau$=1 curve derived using least squares fitting (Table \ref{tbl_S_0}). The surface geometries obtained using Equation (10) match those characterized by the $\tau$=0.5/1/2 curves well for all cases if we use the $PI$ images at a viewing angle of 18$^\circ$. These also match well in cases where we extract the $PI$ distribution at P.A.=90$^\circ$ from the images at a viewing angle of 49$^\circ$. The standard deviation in $z$ depends on the P.A. and the viewing angle, and these at $r > 20$ AU are 0.1--2, 2--5, and 5--8 \% (or 0\farcs14) for Models 1, 2, 3, respectively. We find larger errors for the inner radii probably due to the coarse pixel sampling over the region. Note that in HiCIAO observations this region is either occulted by the coronagraph mask, or not used due to contaminating emission from the star (Section 4). We also find minor discrepancies between between the calculated surface and the $\tau$=0.5/1/2 curves near the disk edge for Models 2 and 3, P.A.= 135$^\circ$ and 180$^\circ$. This does not affect our discussion of surface geometries in later sections.

The surface geometries obtained for Models 2 and 3 show discrepancies from the $\tau$=0.5/1/2 curves at $r > 60$ AU if we use the $PI$ image at a viewing angle of 49$^\circ$ and extract the flux distribution at any P.A. but 90$^\circ$. In later sections, we therefore limit our applications for disks with intermediate viewing angles to the surface geometry along the major axis of the $PI$ distribution. The remaining results confirm that the geometrically thin approximation used in Section 2 produce correct surface geometries up to at least $z/r \sim 0.2$.

Table \ref{tbl_C_max} shows the maximum $C$ defined in Equations (7) and (11) for individual cases. While we have used the approximation $C \ll 1$ to derive Equation (10), we find that Equation (10) reproduces the surface geometry well even for those cases where $C \sim 1$.

\section{APPLICATIONS TO THE SEEDS DATA}

We derive surface geometries for the SEEDS sample usingÊ the differential equation developed in Section 2. 
In Section 4.1 we summarize the objects and parameters used for our analysis. In Section 4.2 we show surface geometries corresponding to maxima in the $PI$ distribution along the radial direction. In Section 4.3 we show an extensive analysis of surface geometries in the azimuthal direction. In Section 4.4 we verify our results by reproducing the observed $PI$ flux distributions for some objects using Monte-Carlo scattering simulations.

\subsection{Objects and Parameters}

Table \ref{tbl_objects} shows the list of objects we have analyzed. This includes disks with spiral structures (SAO 206462, MWC 758), ring-like morphologies (PDS 70, 2MASS J16042165--2130284), and a relatively uniform distribution (MWC 480). The initial SEEDS results for these objects have been published by \citet{Muto12,Hashimoto12,Kusakabe12,Mayama12,Grady13}. All except PDS 70 show bright millimeter/sub-millimeter emission \citep{Hamidouche06,Brown09,Isella10b,Andrews11,Mathews12}, indicating the youth of these disks, and therefore suggesting an optically thick nature in the near-infrared (Section 2).
A 0\farcs3 diameter circular occulting mask was used to suppress the bright stellar flux for SAO 206462, MWC 758 and MWC 480. The occulting mask was not used for the observations of PDS 70 or 2MASS J1604. See the above papers for the details of data reduction.

We need to scale the observed $PI$ flux by the spatially integrated stellar $I$ flux to apply Equation (10). The stellar $I$ flux for PDS 70 was measured using the same data set as those used for the $PI$ observations. For the other objects we used short exposures with a neutral density filter without a coronagraphic mask.
These images were taken on the same date as the coronagraphic images.
See \citet{Takami13} for technical details of the measurement.

We use the same dust model used in Section 3. As shown in Table \ref{tbl_PI_I0} this model yields $(PI/I_0)=$0.012 sr$^{-1}$ with a scattering angle of $90^\circ$ \citep[][]{Takami13}.
This is the approximate scattering angle expected along the major axis of the $PI$ flux distribution, or the entire disk for a face-on view.
Dust grains may be smaller than those defined by the KMH model in some disks (Appendix A). As seen in Equation (10), the use of different dust models (and the corresponding $PI/I_0$) results in the linear scaling of the surface geometry derived. This scaling does not affect the main conclusions of the paper.

To derive the surface geometry several P.A.s were initially chosen for SAO 206462, MWC 758, and 2MASS J1604, which clearly show non-axisymmetric distributions\footnote{\citet{Mayama12} reported the presence of an arc-like structure within the 2MASS J1604 ring at the west side of the star. This structure is now suspected to be an artifact, and is therefore not included in our analysis.} (Figures \ref{fig_SAO}--\ref{fig_J1604}). These disks have inclination angles of $11^\circ$, $21^\circ$, and $10^\circ$, respectively, implying that the scattering angles are slightly different at different P.A.s. Thus, the true $(PI/I_0)$ should be slightly smaller in practice than that determined at 90$^\circ$, depending on the position. These differences are 6--7 \% for SAO 206462 and 2MASS J1604, and 17 \% for MWC 758 along the minor axis, where we expect the largest deviation, assuming the same scattering properties of dust grains measured for PDS 70 and MWC 480 (Appendix A). For simplicity we use the ($PI/I_0$) derived for scattering angles of $90^\circ$ over the entire disk. The variations in the corrections for different scattering angles are significantly smaller than the calculated variation in the disk surface structure or illumination shown below, therefore this approximation does not affect our conclusions.
The PDS 70 and MWC 480 disks have fairly large inclination angles, therefore we derive the surface geometries along the major axis only (see Section 3). We calculate the constant $C$ defined in Equations (7) and (11) to be $\sim$0.3 or smaller for all the objects and P.A.s described above. This agrees with the approximation we have used to derive Equation (10) ($C \ll 1$).

The numerical integrations for 2MASS J1604 and PDS 70 were made starting at  0\farcs25  ($\sim$35 AU) from the star, approximately corresponding to the minimum in the radial $PI$ flux in the ring structure. The integrations for the other objects were made starting from 0\farcs2 (28 AU for SAO 206462 and MWC 480; 40 AU for MWC 758) from the star, i.e., slightly outside the coronagraphic mask and the innermost radius at which we have reliable measurements of the $PI$ flux distribution.

%
%
%
\subsection{Surface Geometries Associated with $PI$ Maxima}

\begin{figure*}
\epsscale{1.8}
\vspace{-2.5cm}
\plotone{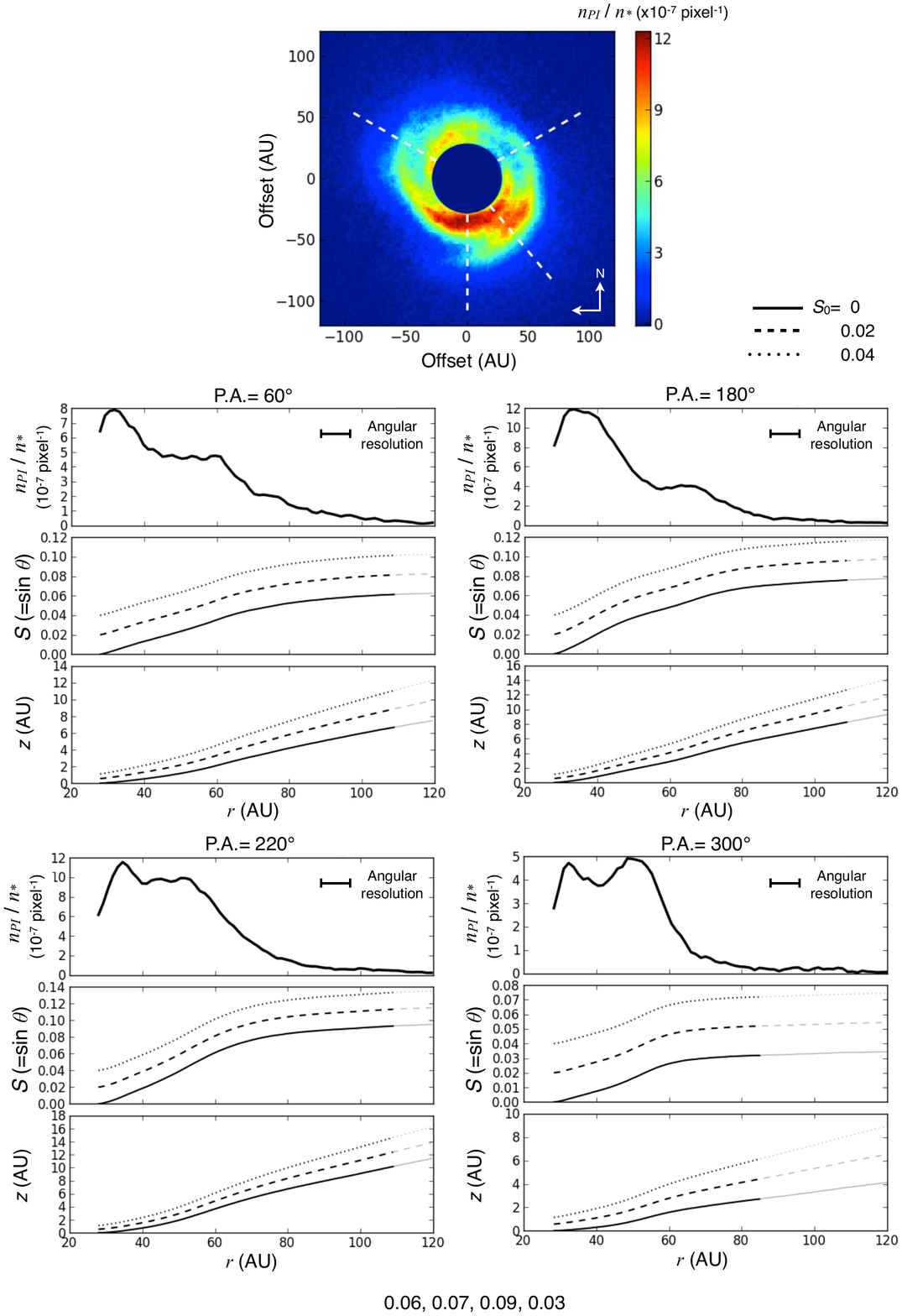}
\caption{
\footnotesize
($top$) $PI$ image of SAO 206462 \citep{Muto12}. The angular resolution is 0\farcs06, corresponding to 8 AU in the image. The $PI$ flux is scaled by the spatially integrated stellar $I$ flux (see text for details). The region within 0\farcs2 (28 AU) of the star is masked. The white dashed lines indicate the positions for which we derive the surface geometry. ($bottom$) The results of the numerical integration for the individual P.A.s. For each P.A., the top figure shows the $PI$ flux distribution; the middle shows $S$=sin $\theta$, where $\theta$ is the elevation angle of the position of a scattering event from the midplane (see Figure \ref{fig_geometry}); the bottom shows the disk surface. The distance $r$ is corrected for the disk inclination.
The initial values of the numerical integration $S_0$ are set to be 0, 0.02, and 0.04 at 0\farcs2 (corresponding to the projected radius of 28 AU) from the star, and the results are shown with solid, dashed, and dotted curves, respectively. Fainter curves at large radii show the results in the outer regions where $n_{PI}/n_* \sim 0$, and the results are therefore not reliable (see text for details).
\label{fig_SAO}}
\end{figure*}

\begin{figure*}
\epsscale{1.8}
\vspace{-2cm}
\plotone{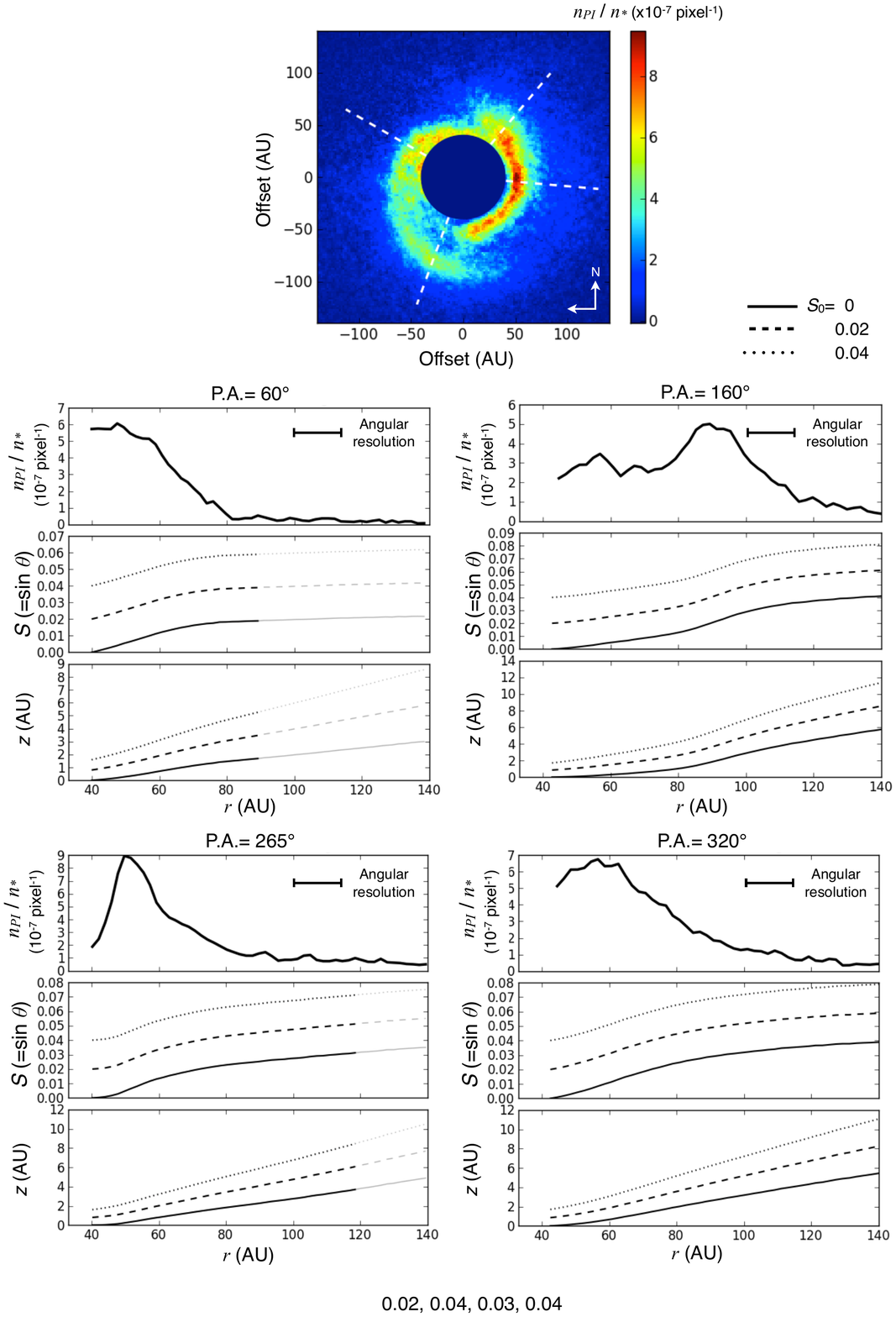}
\caption{Same as Figure \ref{fig_SAO} but for MWC 758 \citep{Grady13}. The angular resolution of 0\farcs07 corresponds to 14 AU. The initial values of the numerical integration $S_0$ (= sin $\theta_0$, where $\theta_0$ is the elevation angle of the position of a scattering event from the midplane) are set to be 0, 0.02, and 0.04 at 0\farcs2 (corresponding to the projected radius of 40 AU) from the star.
\label{fig_MWC758}}
\end{figure*}

\begin{figure*}
\epsscale{1.8}
\vspace{-2cm}
\plotone{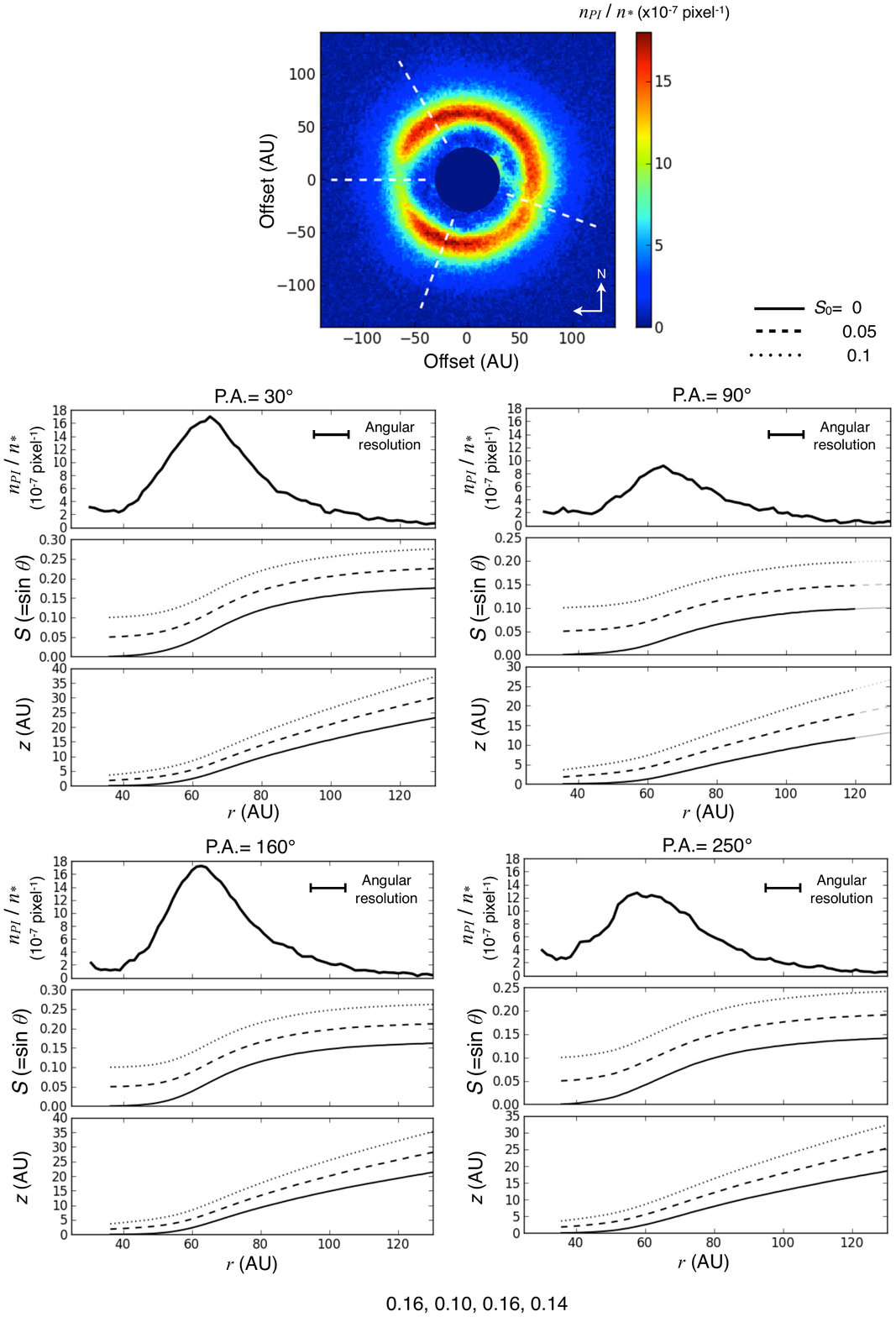}
\caption{Same as Figures \ref{fig_SAO} and \ref{fig_MWC758} but for 2MASS J16042165--2130284 \citep{Mayama12}. The angular resolution of 0\farcs07 corresponds to 10 AU. While the coronagraphic mask is not used, the center of the image is masked due to an artifact caused by the central star. The initial values of the numerical integration $S_0$ (= sin $\theta_0$, where $\theta_0$ is the elevation angle of the position of a scattering event from the midplane) are set to be 0, 0.05, and 0.1 at 0\farcs25 (corresponding to the projected radius of 36 AU) from the star.
\label{fig_J1604}}
\end{figure*}

\begin{figure*}
\epsscale{1.8}
\plotone{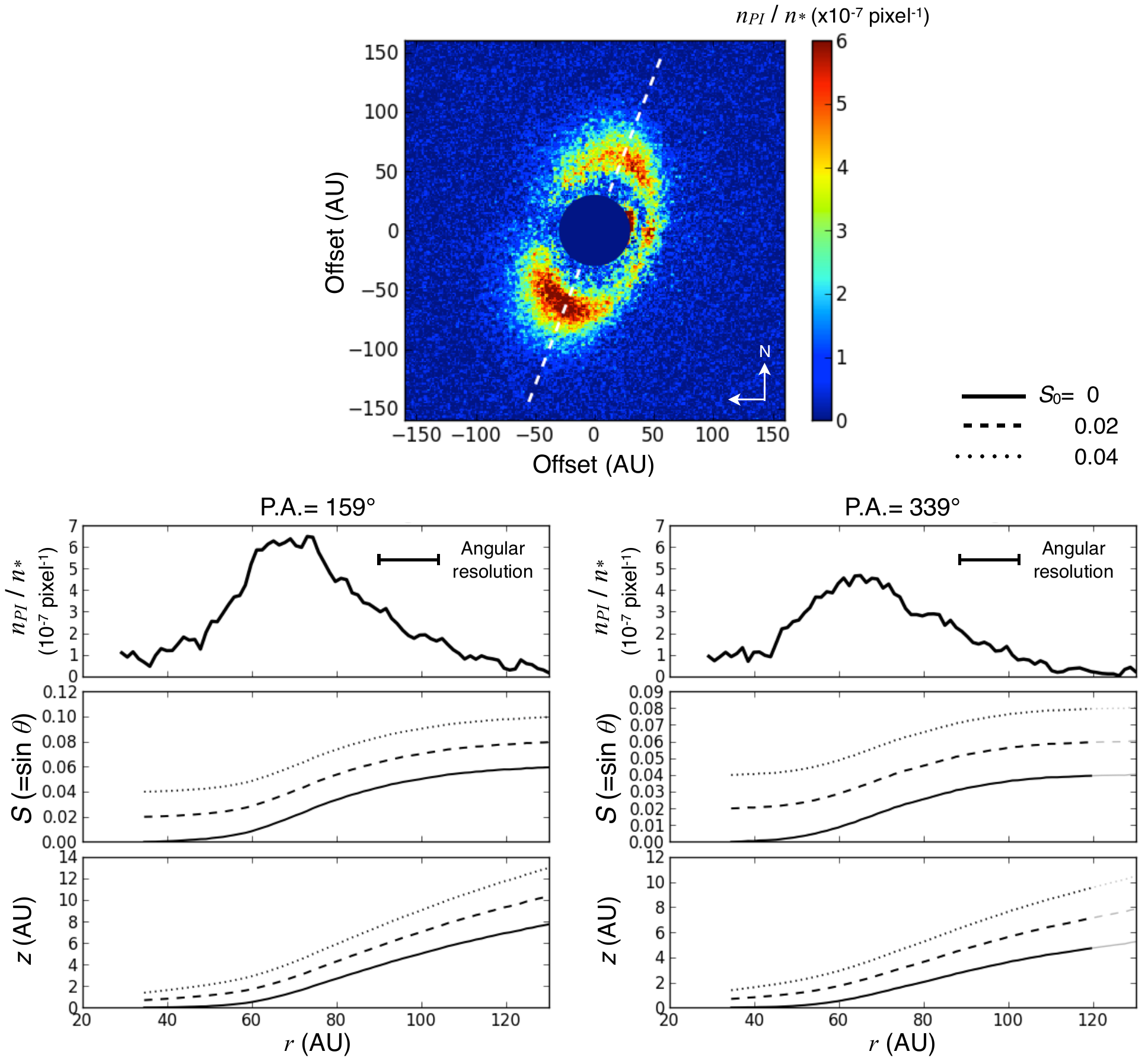}
\caption{Same as Figure \ref{fig_J1604} but for PDS 70 \citep{Hashimoto12}. The angular resolution of 0\farcs1 corresponds to 14 AU. The numerical integrations are made along the major axis of the disk, hence the distance $r$ in the lower plots is equal to the projected distance in the plane of the sky. The initial values of the numerical integration $S_0$ (= sin $\theta_0$, where $\theta_0$ is the elevation angle of the position of a scattering event from the midplane) are set to be 0, 0.02, and 0.04 at 0\farcs25 (corresponding to the projected radius of 35 AU) from the star.
\label{fig_PDS70}}
\end{figure*}

\begin{figure*}
\epsscale{1.8}
\plotone{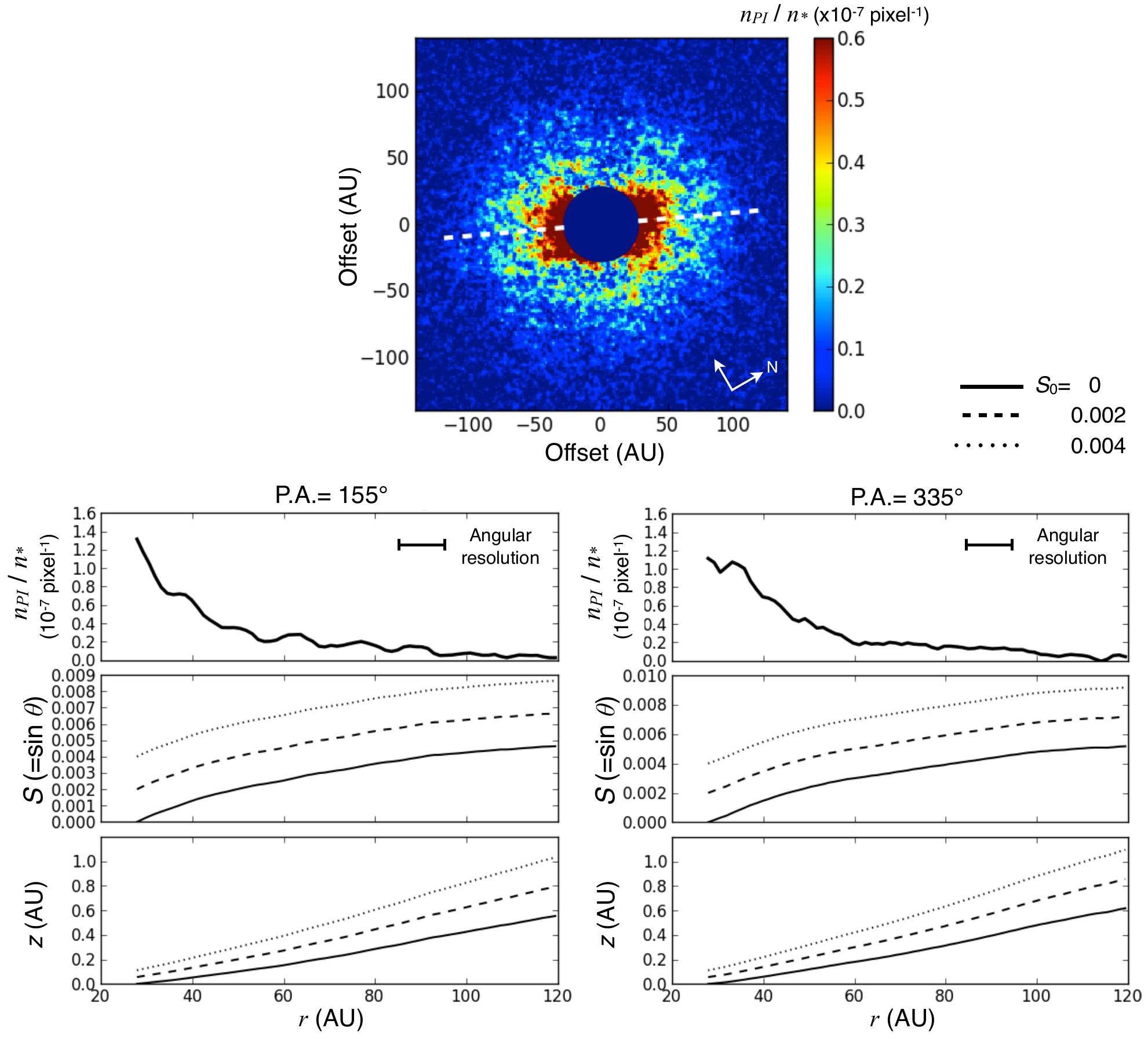}
\caption{Same as Figure \ref{fig_PDS70} but for MWC 480 \citep{Kusakabe12}. The angular resolution of 0\farcs07 corresponds to 10 AU. The initial values of the numerical integration $S_0$ (= sin $\theta_0$, where $\theta_0$ is the elevation angle of the position of a scattering event from the midplane) are set to be 0, 0.002, and 0.004 at 0\farcs2 (corresponding to the projected radius of 28 AU) from the star.
\label{fig_MWC480}}
\end{figure*}

Figures \ref{fig_SAO}--\ref{fig_MWC480} show the observed $PI$ images, the position at which the $PI$ intensity profile was extracted, and the calculated surface geometry. For each object a few values are arbitrarily chosen for $S_0$ (= sin $\theta_0$, where $\theta_0$ is the elevation angle of the position of a scattering event from the midplane at the radius $r_0$) to show how the location of the disk surface varies with this parameter.
The outermost regions with constant or nearly constant $S$ are shown with fainter curves, since we cannot determine if disk material is present or the region is shadowed (Section 2). We  note that the diffuse emission at these radii may also suffer from an artificial halo, a result of the faint halo of the point-spread function smearing the emission in the bright inner regions. In all the plots the height of the disk surface from the midplane is significantly smaller than the corresponding radius, consistent with the first approximation of Section 2. 

In Figures \ref{fig_SAO}--\ref{fig_MWC480} all the objects except MWC 480 show local maxima in the plotted one-dimensional $PI$ distributions. However, none of them show local maxima in the disk height $z$ in these figures. Instead, the parameter $S$ (therefore $\theta$) shows the largest increases near the bright $PI$ maxima, slightly offset toward smaller $r$. These positions are:-
$r \sim 50$ AU, P.A.=300$^\circ$ in SAO 206462;
$r \sim 90$ AU, P.A.=160$^\circ$ in MWC 758; and
the position of the ring in the $PI$ distribution in 2MASS J1604 and PDS 70. For the rest of the paper we describe such geometries as ``local concave-up structures" at the disk surface.
The presence of local concave-up structures is independent of the assumed $S_0$ (=sin $\theta_0$).

As shown in Figures \ref{fig_J1604} and \ref{fig_PDS70} and explained in Section 2, a strong deficit in the $PI$ flux in the ring structure does not imply the absence of an inner disk. This can be explained with either a geometrically thin disk whose surface is nearly parallel to the light path of the star (i.e., a nearly constant $S$ and $\theta$), or self-shadowing by the inner disk (see Section 2). The disk geometries shown in Figures \ref{fig_J1604} and \ref{fig_PDS70} do not show the presence of a sharp vertical inner wall. This contrasts with discussions or assumptions in many observational studies of disks with a hole or gap \citep[see][for a review]{Espaillat14}. We will further explore this issue in Section 4.4 with a monochromatic radiative transfer simulation.

\subsection{Azimuthal Structures}
\begin{figure*}
\epsscale{2}
\vspace{-2.5cm}
\plotone{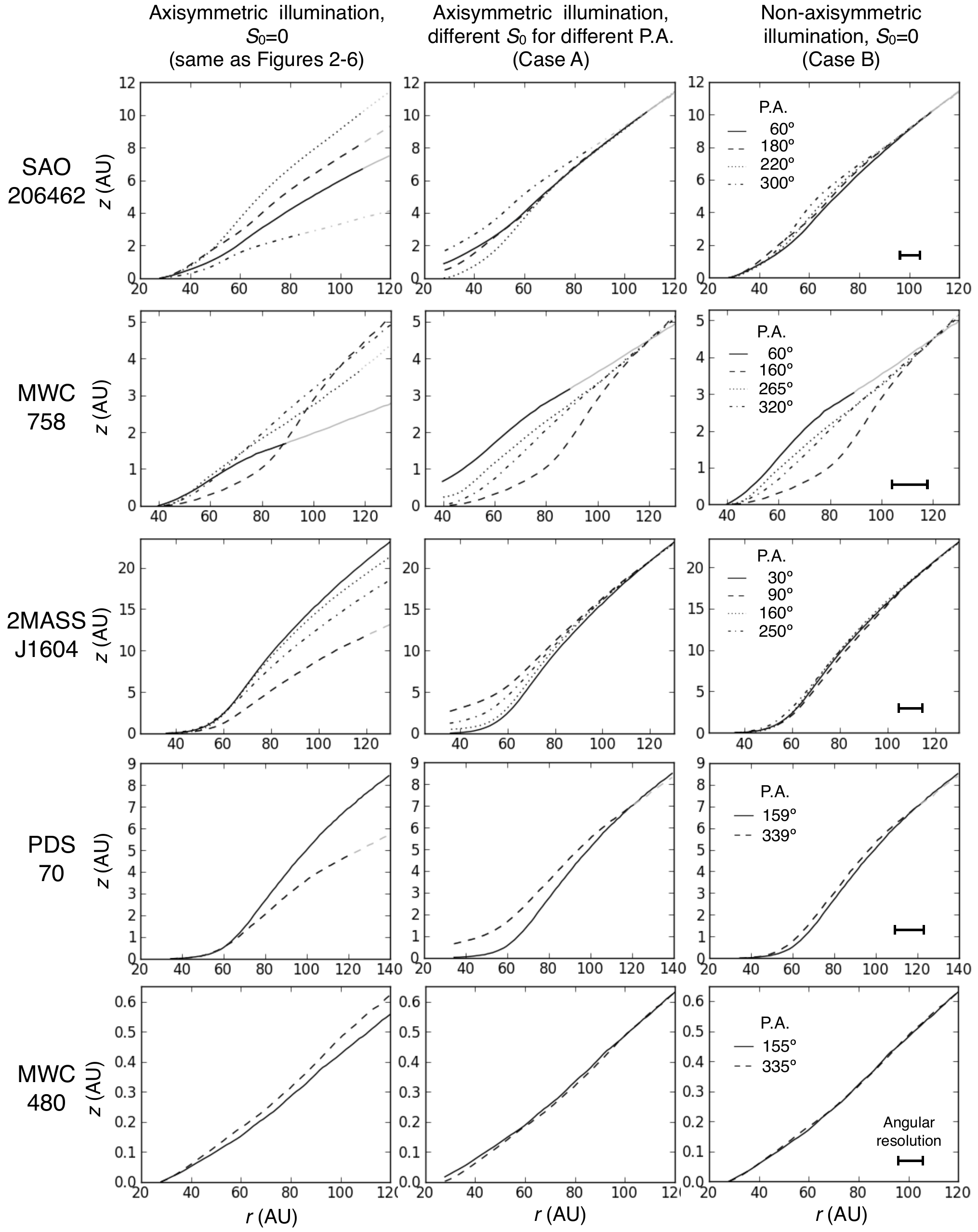}
\caption{Disk surfaces obtained for the following cases: (1) axisymmetric illumination, $S_0$=sin $\theta_0$=0, where $\theta_0$ is the elevation angle of the position of a scattering event from the midplane at the innermost radius; (2) axisymmetric illumination, different $S_0$ for different P.A.s to produce the axisymmetric structure at large radii (Case A); (3) $S_0$=0, non-axisymmetric illumination to produce the axisymmetric structure at large radii (Case B). See text for details. The P.A.s selected are the same as in Figures \ref{fig_SAO}--\ref{fig_MWC480}. As in Figures \ref{fig_SAO}--\ref{fig_MWC480} the fainter curves show the results in the outer regions where the observed $PI$ flux $n_{PI}/n_* \sim 0$, and therefore the results are not reliable.
 \label{fig_revised_z}}
\end{figure*}

We plot the surface geometry for individual objects for $S_0$=0 in the left column of Figure \ref{fig_revised_z}. All the disks except MWC 480 show large non-axisymmetries. This contrast to the axisymmetric geometry which is assumed for many disk models \citep[see e.g.,][for a review]{Dullemond07_PPV}.
We investigate below if a more axisymmetric geometry is possible assuming either: (A) different values of $S_0$ (= sin $\theta_{0}$, where $\theta_{0}$ is the elevation angle of the position of a scattering event from the midplane at distance $r_0$) at different azimuthal angles in the disk coordinate $\phi$ (or the observed P.A.); or (B) a non-axisymmetric illumination of the disk (see Section 5.2 for possible origins of such illumination). In this subsection we will also discuss other behavior in the concave-up and concave-down structures not clearly seen in Figures \ref{fig_SAO}--\ref{fig_MWC480}.

For Case (A) the parameter $S_0$ is obtained as follows. From Equation (10) we derive:
\begin{equation}
S (r_{1}, \phi) = S_0 (\phi) +  \frac{\rm ~cos \it ~i }{n_{\nu;*}  A_{pix}} \left( \frac{PI}{I_{\rm 0}} \right)^{-1}   \int_{r_0}^{r_{1}} r n_{\nu;PI}(r,\phi) dr, 
\end{equation}
where $r_{1}$ is the outer radius;
$\phi$ is the azimuthal angle in the disk plane;
$i$ is the viewing angle of the disk;
$n_*$ is the observed stellar $I$ flux;
$A_{pix}$ is the area corresponding to the pixel scale of the instrument;
($PI/I_0$) is the $PI$ flux normalized to the incident $I$ flux determined by the dust properties;
$r_0$ is the inner radius where we define $S_0$;
$n_{PI}$ is the observed $PI$ flux distribution from the disk. 
 If the disk height is constant at the radius $r_1$,
$S(r_{1}, \phi)= S_{1}$ (i.e., a constant). Substituting this into Equation (13) we derive $S_0 (\phi)$ and a disk height $z (\phi)$ of:
\begin{eqnarray}
S_0 (\phi) & = & S_1 - \frac{\rm ~cos \it ~i }{n_{\nu;*}  A_{pix}} \left( \frac{PI}{I_{\rm 0}} \right)^{-1}  \int_{r_0}^{r_{1}} r n_{\nu;PI}(r,\phi) dr \nonumber \\
 & = & S_1 - S(r_{1},\phi, S_0=0).
\end{eqnarray}
For Case (B) Equation (10) becomes:
\begin{equation}
dS  =  r \frac{n_{\nu;PI}(r, \phi)}{f(\phi) n_{\nu;*} }  \frac{\rm ~cos \it ~i }{A_{pix}} \left( \frac{PI}{I_{\rm 0}} \right)^{-1} dr, 
\end{equation}
where $f(\phi)$ is the fraction of stellar flux passing through the dust screen between the star and the disk surface. Thus,
\begin{equation}
S(r_{1}, \phi)  =  \frac{\rm ~cos \it ~i }{f(\phi)  n_{\nu;*}  A_{pix}}  \left( \frac{PI}{I_{\rm 0}} \right)^{-1}   \int_{r_0}^{r_{1}} r n_{\nu;PI}(r,\phi) dr, 
\end{equation}
if $S_0$ = sin $\theta_0$ = 0. Substituting $S (r_{1}, \phi)= S_1$ we derive:
\begin{eqnarray}
f(\phi) & = & \left[ \frac{\rm ~cos \it ~i }{n_{\nu;*}  A_{pix}} \left( \frac{PI}{I_{\rm 0}} \right)^{-1}  \int_{r_0}^{r_{1}} r n_{\nu;PI}(r,\phi) dr \right] / S_1 \nonumber \\
& = & \frac{S (r_{1}, \phi, f=1, S_0=0)}{S_1}.
\end{eqnarray}

In Figure \ref{fig_revised_z} we plot the revised locations of the disk surface for Cases (A) (i.e., different $S_0$ for different P.A.s) and (B) (non-axisymmetric illumination, $S_0$=0).
Table \ref{tbl_params_for_revised_z} shows the parameters used for Cases (A) and (B) in Figure \ref{fig_revised_z}.
The parameter $S_1$ at radius $r_1$ is chosen to be an approximately minimum possible value for each object, which results in the minimum $S_0$ and the maximum $f$ being $\sim 0$ and $\sim 1$, respectively. 
The use of different $S_1$ values does not significantly change the results described below.

In Figure \ref{fig_revised_z} the surface geometries at different P.A.s are similar, in particular for Case (B) (i.e., with non-axisymmetric illumination) with the exception of the MWC 758 disk, P.A.=160$^\circ$. The MWC758 disk shows a clear concave-up geometry at $r$=60--100 AU at this position angle. This concave-up geometry is associated with a faint arm in the southeast, seen in the $PI$ image in Figure \ref{fig_MWC758}.

Figure \ref{fig_revised_z} shows that Equations (8) and (11) produce the same local concave-up and concave-down structures. 
The surface geometry is close to a straight line for all the other P.A.s for SAO 206462 and MWC 758 which are associated with spiral structures. In other words, most of the spiral structures in these objects are attributed to surface structures whose spatial variation in the $z$ direction is significantly smaller than the thickness of the disk.
As shown in Figures \ref{fig_J1604}--\ref{fig_MWC480}, the disk surfaces in 2MASS J1604 and PDS 70 show a distinct concave-up geometry at small radii, and a marginal concave-down geometry at large radii. 
The surface in MWC 480 shows a marginal concave-up geometry.

\begin{figure*}
\epsscale{2.1}
\plotone{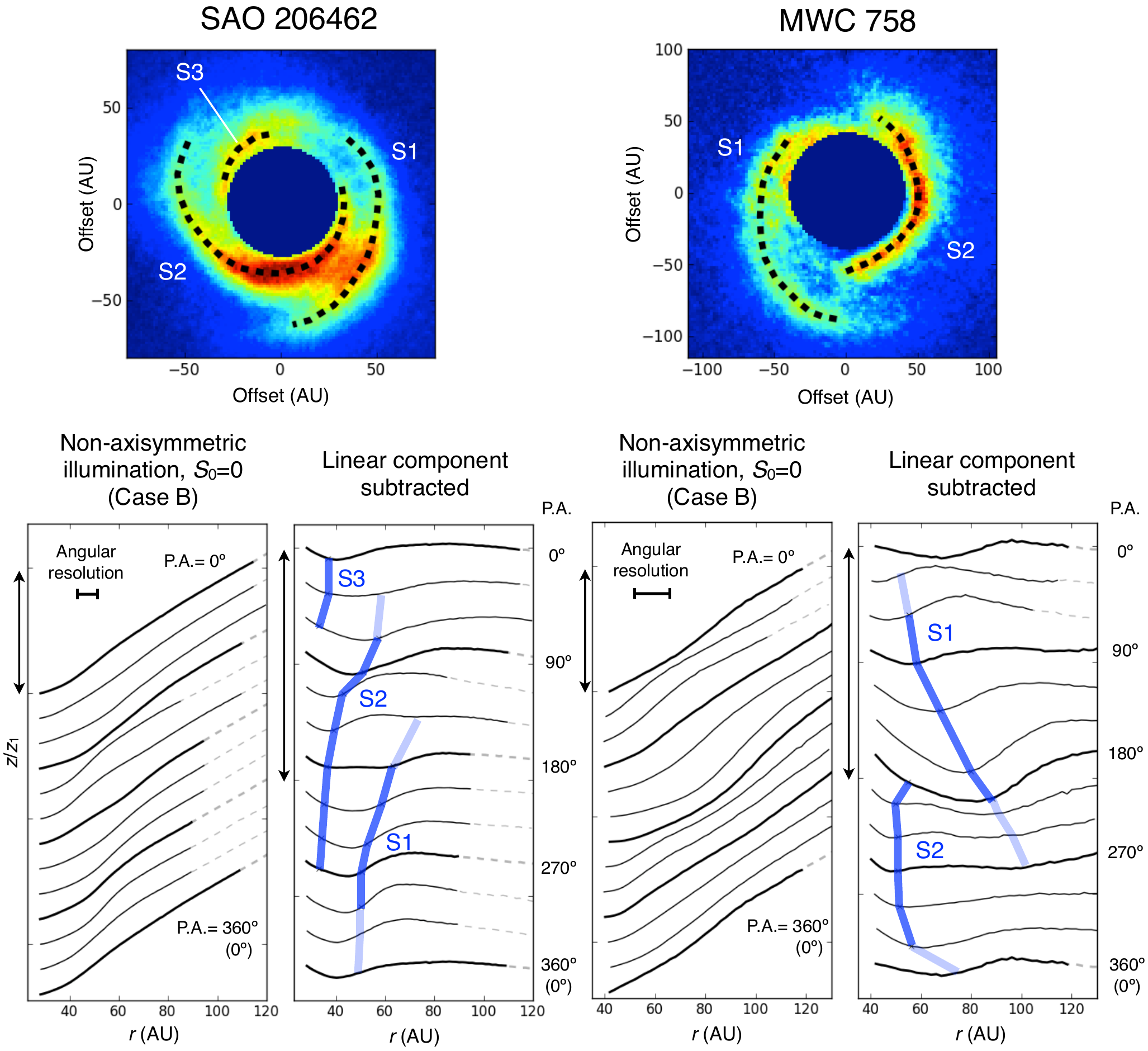}
\caption{($top$) Spiral arms discussed for SAO 206462 and MWC 758. ($bottom$) Surface geometry for Case (B) at different P.A.s (left for each object) and those for which we subtracted the linear component (right, see the text for details). These are obtained at P.A.s with an interval of 30$^\circ$, and each of them is shown with an arbitrary offset. The vertical arrows correspond to $z/z_1=1$, where $z_1$ is the disk height at the outermost radius $r_1$ (see Table  \ref{tbl_params_for_revised_z}). The thick curves show the geometries for P.A.=0$^\circ$, 90$^\circ$, 180$^\circ$, and 270$^\circ$.  Blue curves show the positions of the spiral arms indicated by black dashed curves in the upper figures. The faint blue curves at the ends of some of the blue curves are the positions of arms not clearly seen in the top image, but identified as local concave-up structures at the disk surface.
\label{fig_revised_z_every_30deg}}
\end{figure*}

What is the relationship between the spiral structures and the local concave-up/down surface geometries in the SAO 206462 and MWC 758 disks? To clearly demonstrate this, we plot the disk height in these objects with P.A.s at 30$^\circ$ intervals for Case (B) (non-axisymmetric illumination, $S_0$=0) in Figure \ref{fig_revised_z_every_30deg}. To easily separate concave-up/down structures from the effects of the surface inclination of the disk, we also subtract the linear component of the surface geometry characterized by the positions ($r_0$,0) and ($r_1$, $z_1$), i.e., $z=z_0(r-r_0)/(r_1-r_0)$, and plot them in the same figure.

In Figure \ref{fig_revised_z_every_30deg} all the local concave-up structures are attributed to a few prominent spiral arms identified in the $PI$ distribution. There may also be a concave-up structure (thereby a spiral arm) at P.A.=30$^\circ$--90$^\circ$, $r$=100--120 AU in the MWC 758 disk, but a higher signal-to-noise in the $PI$ flux is required to confirm this.



\subsection{Verification of the Derived Surface Geometries Using Scattering Simulations}

As discussed in the above subsections, we demonstrate that our approach is very useful for determining the surface geometry from the observed $PI$ distribution. In order to reinforce our approach and results, we will now reproduce the observed $PI$ flux distribution for PDS 70 and SAO 206462 using the disk geometry computed by equation (10) and Monte-Carlo scattering simulations as described below.

As for Section 3 a central unpolarized point source (star) serves as the starting point for calculating the scattering of photons from the disk surface. We use the same interstellar dust model as in Section 3. For each photon reaching the disk surface, we calculate the scattering angle and Stokes parameters based on Mie theory. The Stokes parameters for each photon are initially set to ($I_0$,$Q_0$,$U_0$,$V_0$)=(1,0,0,0) and normalized to $\it I_{\rm out} \it = albedo \cdot \it I_{\rm in}$ after each scattering. The photons escaping from the disk are collected in a imaginary detector at a viewing angle of the observations. As outlined in Section 2, multiple scatterings are not included, and photons scattered inward to the disk are discarded. 

For PDS 70 we use an axisymmetric surface geometry parameterized by $S$ (=sin $\theta$, where $\theta$ is the elevation angle of the position of a scattering event from the midplane) as for Section 4.2. The centroid of the observed ring structure is offset by $\sim$6 AU along the minor axis due to its thickness in the vertical direction \citep{Hashimoto12}. The parameter $S_0$ is set to 0.1 at $r_0$=35 AU to reproduce this offset. We derive the height of the surface $z$ by averaging the values at two P.A.s along the major axis (159$^\circ$ and 339$^\circ$). This surface geometry produces disk heights of 3.5 and 21 AU (5$^\circ$.7 and 8$^\circ$.5 from the midplane) at $r$=35 and 140 AU, respectively. The viewing angle of the simulation is set to be 50$^\circ$, matching the observed image, integrated over a range of $\pm 10^\circ$.
We do not include illumination below 5$^\circ$.7 from the midplane, which may cause back scattering at the inner edge of the ring. To produce a sufficient signal-to-noise, we used $10^7$ photons for the model. In order to normalize the $PI$ flux to the stellar $I$ flux, we separately calculate the expected number of photons based on the incident number of photons.

\begin{figure*}
\epsscale{2.2}
\plotone{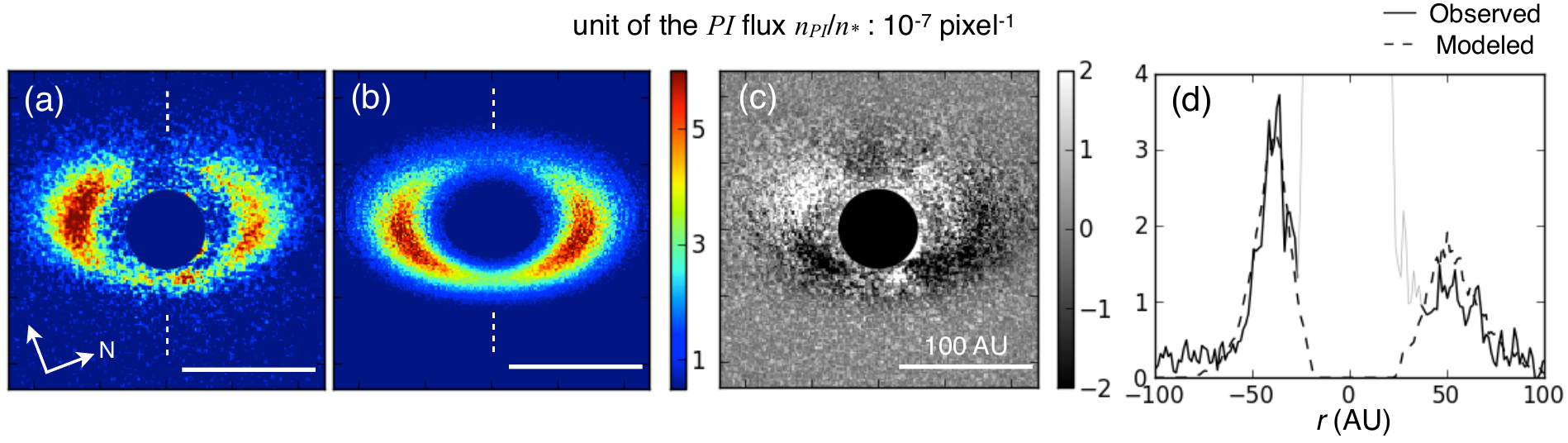}
\caption{$(a)(b)$ Observed and modeled $PI$ flux distributions for PDS 70. The solid lines show a spatial scale of 100 AU. Dashed lines show the minor axis of the ring, where we extract the one-dimensional distribution.
$(c)$ Residual of the modeled image subtracted from the observed image. $(d)$ Observed and modeled $PI$ flux distributions along the minor axis. For each image the flux was averaged across the minor axis with a 0\farcs1 bin (corresponding to 14 AU). The fainter curves show the artifact due to the star, which is masked in (a) and (c). For $(a)(b)(c)(d)$ the $PI$ fluxes $n_{PI}/n_*$ are shown in units of $10^{-7}$ pixel$^{-1}$.
\label{fig_PDS70_obs_vs_model}}
\end{figure*}

Figure \ref{fig_PDS70_obs_vs_model} shows the comparison between the observed and modeled $PI$ distributions. The figure shows that the modeled $PI$ distribution explains the observed $PI$ distribution fairly well, typically within 30 \% in flux. The observed and modeled $PI$ distributions match particularly well along the minor axis, where one might expect back scattering from the vertical wall at the far side of the ring. Figure \ref{fig_PDS70_obs_vs_model} (c) shows the residuals of the modeled image subtracted from the observed image. The observed fluxes are marginally larger than the model fluxes in the southwest of the ring, and marginally lower in the northeast of the ring. These may be due to either of the following: (1) the size distribution (and thereby the scattering properties) of dust grains deviating from the dust model  we have used (Appendix A), or (2) local deviation from  symmetry of the surface geometry.

\begin{figure*}
\epsscale{1.6}
\plotone{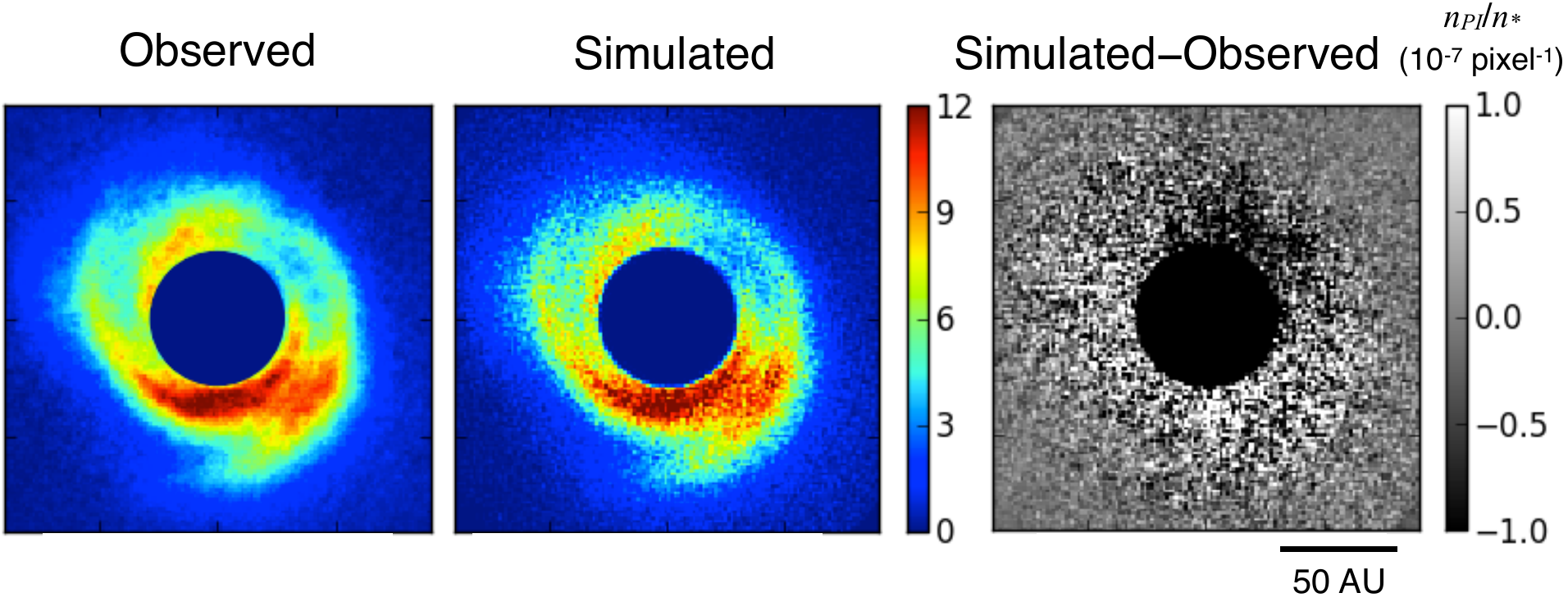}
\caption{$(a)(b)$ Observed and simulated $PI$ flux distributions for SAO 206462. The residual between the observed and simulated images is also shown. See text for details of the simulation.
\label{fig_SAO_obs_vs_model}}
\end{figure*}

SAO 206462 shows a non-asisymmetric $PI$ distribution a close to face-on view. We therefore derived the disk surface for 36 P.A.s at $10^\circ$ intervals from the observed $PI$ image for $S_0$=0. The surface positions between these P.A.s are linearly interpolated. The imaginary detector is set to collect photons at a P.A. of $155^\circ$ integrated over a range of $\pm10^\circ$ and a viewing angle of $11^\circ$ integrated over a range of $\pm5^\circ$ in order to match the observed image. We have used 10$^9$ photons for the simulated image. Figure \ref{fig_SAO_obs_vs_model} shows the observed and simulated images, which match well.



\section{DISCUSSION}

In Sections 4 we have obtained the surface geometry of five disks using the SEEDS data. The major results are summarized below:
\begin{itemize}
\item Spiral and ring structures observed in the $PI$ distributions are associated with local concave-up geometries at the disk surface;
\item Most of the concave-up geometries associated with spiral structures have vertical spatial amplitudes significantly smaller than the thickness the disk;
\item The surface geometries obtained for the ring structures do not show clear evidence for an inner vertical wall like those discussed in the literature for many disks with a hole or gap. The deficit of flux in the ring may be attributed to a disk surface nearly parallel to the light path from the star, or self-shadowing of the region by the inner disk;
\item A combination of relatively axisymmetric geometries and non-axisymmetric illumination can explain the observed $PI$ distribution for most of the disks.
\end{itemize}

In Section 5.1 we discuss possible implications for the surface geometries associated with spiral and ring structures. In Section 5.2 we discuss possible mechanisms for non-axisymmetric illumination of the disk.

\subsection{Spiral and Ring Structures in the Near Infrared}

Young planets may tidally interact with protoplanetary disks, resulting in spiral structures in the disk \citep[e.g.,][for a review]{Papaloizou_PPV}. For example, the spiral structures observed in the SAO 206462 and MWC 758 disks have been discussed as a possible signpost of ongoing planet formation \citep{Muto12,Grady13}. 

Tidal interaction with planets may cause a variety of waves in protoplanetary disks. \citet{Lubow98} categorized as follows based on analogy with the theory of stellar oscillations:
(1) $p$-modes, i.e., acoustic modes responsible for global spiral density waves;
(2) $f$-modes, which are surface gravity waves;
(3) $g$-modes, gravity modes relying on buoyancy forces resulting from an entropy gradient in the vertical direction; and
(4) $r$-modes, which propagate by inertial forces.
Spiral structures in the disk have been investigated mainly in the surface density distribution \citep[e.g.,][for a review]{Papaloizou_PPV}, where they are incited by $p$-modes. In contrast, $f$- and $g$-modes may also create near-infrared observable spiral structures at the disk surface, but they do not significantly affect the surface density distribution of the disk.
In particular, \citet{Lubow98} suggest that the $f$-modes carry most of the torque exerted at the resonance. These modes may induce spiral shock waves observable at the surface of the disk \citep{Boley06}. \citet{Zhu12b} discussed the importance of $g$-modes in disks based on their simulations which show  3-D structures of  spiral waves induced by tidal interaction with a young planet.

Comparisons between near-infrared and millimeter/sub-millimeter images would give useful constraints for understanding which modes are responsible for spiral arms, and thereby understanding the internal structures in the disk in a manner similar to stellar seismology. The millimeter/sub-millimeter dust continuum emission from the disk is optically thin \citep[e.g.,][for reviews]{Williams11,Espaillat14}, therefore the emission would trace the surface density distribution.
Sub-mm images for SAO 206462 and MWC 758 have been obtained by \citet{Brown09,Isella10b,Andrews11} with angular resolutions of 0\farcs3--0\farcs8 (corresponding to 42--110 AU and 60--160 AU, respectively). These images do not show the presence of spiral structures expected for $p$-modes. However, these modes cannot be fully ruled out without millimeter/sub-millimeter images at angular resolutions similar to our near-infrared images ($0\farcs05-0\farcs1$).


Disk holes have been directly observed in dozens of protoplanetary disks, including those in the 2MASS J1604 and PDS 70 rings, with near-infrared and millimeter/sub-millimeter imaging techniques. Such structures have also been inferred from defecits at certain wavelengths in infrared SEDs \citep[for reviews]{Williams11,Espaillat14}. As with spiral structures, these holes have been discussed as potential signatures of ongoing planet planet formation and tidal interaction between the disk and planet(s). Such disks, which have been termed ``Transitional disks", are often assumed to be associated with a vertical wall at the edge of the hole or the gap \citep[][for a review]{Espaillat14}.
Scattered light at optical and near-infrared wavelengths agrees with the presence of such a wall in GG Tau \citep{Silber00,Duchene04} and LkCa 15 \citep{Thalmann10}.
Such a geometry is consistent with 3-D simulations by \citet{Zhu12a} for disk clearing by planetary-mass companions without dust filteration (i.e., without dust and gas accreting through the gap).

In contrast, our analysis of the 2MASS J1604 and PDS 70 rings shows that the observed near-infrared $PI$ distributions can be explained without back scattering at the inner wall. The surface geometries in Figures \ref{fig_J1604}, \ref{fig_PDS70}, and \ref{fig_revised_z} do not show a discontinuous boundary at the inner edge of the ring. The disk may still exist at small radii with very low $PI$ fluxes. Millmeter/submillimeter interferometry with high angular resolutions, which would allow us to sufficiently resolve the ring, is necessary to further investigate the nature of ring structures \citep[e.g.,][for a review]{Espaillat14}.


\subsection{Possible Non-axisymmetric Illumination}

The 2MASS J1604 ring shows a remarkable deficit in the $PI$ flux distribution at P.A.$\sim 90^\circ$ and a marginal deficit at P.A.$= 180-270^\circ$ \citep[Figure \ref{fig_J1604}]{Mayama12}. PDS 70 and MWC 480 show asymmetries in the $PI$ distribution with respect to the minor axis  \citep[Figures \ref{fig_PDS70} and \ref{fig_MWC480}]{Hashimoto12,Dong12b,Kusakabe12}. 
Possible explanations for such asymmetries include differences in the scattering geometry, dust mass distribution and dust properties \citep{Mayama12, Dong12b}. \citet{Mayama12} also discussed the possibility that the deficit at P.A.$\sim 90^\circ$ \citep[``Dip {\bf D}'' according to][]{Mayama12} is associated with an embedded protoplanet.

Some results of our analysis in Section 4.2 are consistent with a nearly axisymmetric surface geometry with non-axisymmetric illumination. Such illumination has been discussed for several disks, including those associated with HH 30  \citep[e.g.,][]{Watson07_HH30}, GG Tau \citep{Silber00,Itoh02,Krist05}, HD 163296 \citep{Wisniewski08}, and RY Tau \citep{Takami13}. Possible mechanisms for non-axisymmetric illumination include: (1) hot (or cool) spots on the star; (2) shadowing by a non-axisymmetric inner disk, including a puffed-up inner rim; (3) obscuration by a companion star; and (4) obscuration by a disk associated with a companion star \citep{Watson07_HH30,Wisniewski08,Takami13,Dullemond03}. In addition to these mechanisms, clumpy accretion of dust and gas onto the star or an inner wind associated with the star may also explain the observed asymmetry, as has been proposed for UX-Ori type variables  \citep[see][and references therein]{Herbst94,Grady00}.

Dip {\bf D} in the 2MASS J1604 ring is narrow in P.A., and may be explained by obscuration by a non-axisymmetric distribution of dust in the inner disk or clumpy mass accretion. The same interpretation was made by \citet{Silber00,Itoh02,Krist05} for a similar structure in the scattered light of the GG Tau ring. In contrast, the asymmetric distribution in 2MASS J1604 and MWC 480 may be explained by either hot/cool spots or by the presence of a companion, both of which provide an asymmetry at a wide range of P.A.s. Alternatively, these asymmetried may be caused by self-shadowing by a warped or misaligned inner disk \citep{Whitney13}.

Synoptic observations of the $PI$ distribution would be extremely useful to discriminate between the above mechanisms. For example, a rotation period of up to 10--20 days would be expected if non-axisymmetric illumination is due to the non-uniform surface brightness of the star \citep[][for a review]{Herbst07}. In contrast, a significantly longer period (e.g., 10 years or more) would be expected if the non-axisymmetric illumination is due to shadowing by a non-axisymmetric inner disk \citep{Krist05,Watson07_HH30,Wisniewski08,Takami13}. The rotational period of the disk depends on radius, therefore measurements of the speed of the non-axisymmetric patterns would be useful for identifying the radius of the obscuration.
Synoptic observations would also allow us to further investigate whether the rotation speed of the pattern agrees with predictions of the spiral density wave theory \citep{Muto12}. 

\section{CONCLUSIONS}

We have developed a new method to derive the surface geometry of protoplanetary disks based on observations at near-infrared and optical wavelengths. We have obtained a differential equation to derive the surface geometry from the observed flux distribution, the disk inclination, and assumed properties of dust scattering, with the following approximations: (1) the disk is geometrically thin and optically thick; (2) the width of the scattering layer is negligible; and (3) multiple scattering effects are negligible. Comparing with monochromatic Monte-Carlo radiative transfer simulations, we find that this method is applicable for any P.A. for face-on disks, and along the major axis of the flux distribution for disks with intermediate viewing angles.

We have applied this method to the $PI$ images of five disks using Subaru-HiCIAO at 1.65 \micron. The objects include those with spiral structures (SAO 206462, Muto et al. 2012; MWC 758, Grady et al. 2013), ring structures (2MASS J16042165--2130284, Mayama et al. 2012; PDS 70, Hashimoto et al. 2012), and those with a relatively uniform distribution \citep[MWC 480,][]{Kusakabe12}. We find that the local $PI$ maxima in spiral arms and rings are associated with local concave-up structures at the disk surface if the above approximations are appropriate. The southern arm in the MWC 758 disk is associated with a remarkable concave-up structure despite its relatively faint $PI$ flux. In contrast, the other spiral arms in the SAO 206462 and MWC 758 disks are associated with concave-up geometries whose spatial amplitudes in the vertical direction are significantly smaller than the disk height.

The very low $PI$ fluxes in the 2MASS J16042165--2130284 and PDS 70 rings can be explained by either a geometrically thin disk whose surface is nearly parallel to the light path of the star, or self-shadowing by the inner disk. In other words, the presence of a disk hole in the $PI$ distribution alone does not prove the absence of the disk at the radii of the apparent disk hole. The geometries we obtained do not show evidence for a vertical wall associated with the disk hole. The absence of an inner vertical wall is corroborated by comparisons between the observations and a monochromatic scattering simulation of the PDS 70 ring.

Previous studies suggest that different scattering geometries are responsible for the non-uniform $PI$ distribution in 2MASS J16042165--2130284 and PDS 70. Through our analysis of the surface geometry, we add the possibility that the disk surface is centrosymmetric and the non-uniform $PI$ distribution is attributable to non-axisymmetric illumination due to obscuration by the inner disk, an accretion flow or a wind.

\acknowledgments
We are grateful for anonymous referees for thorough reviews and valuable comments.
We thank the Subaru Telescope staff for their support, especially from Michael Lemmen for making our observations successful.
We thank Dr. Hyosun Kim for useful discussion.
M.T. is supported from Ministry of Science and Technology (MoST) of Taiwan (Grant No. {100-2112-M-001-007-MY3} and 103-2112-M-001-029).
Y.H. is supported by the EACOA Fellowship that is supported by the East Asia Core Observatories Association which consists of the Academia Sinica Institute of Astronomy and Astrophysics, the National Astronomical Observatory of Japan, the National Astronomical Observatory of China, and the Korea Astronomy and Space Science Institute.
T.M. is supported by JSPS KAKENHI Grant Numbers 26800106, 23103004, 26400224.
R.D. acknowledges the support for this work by NASA through Hubble
Fellowship grant HST-HF-51320.01-A awarded by the Space Telescope Science
Institute, which is operated by the Association of Universities for
Research in Astronomy, Inc., for NASA, under contract NAS 5-26555.
Jungmi Kwon is supported by the JSPS Research Fellowships for Young Scientists (PD: 24$\cdot$110).
J.C. was supported by NSF-AST 1009203. 
C.A.G. acknowledges support under NSF AST 1008440.
J.P.W. is supported by NSF-AST 1009314.
This research made use of the Simbad data base operated at CDS, Strasbourg, France, and the NASA's Astrophysics Data System Abstract Service.

{\it Facilities:} \facility{Subaru (HiCIAO)}.

\appendix





\section{Observed and modeled $PI$ fluxes as a function of scattering angle}
\begin{figure*}
\epsscale{1.6}
\plotone{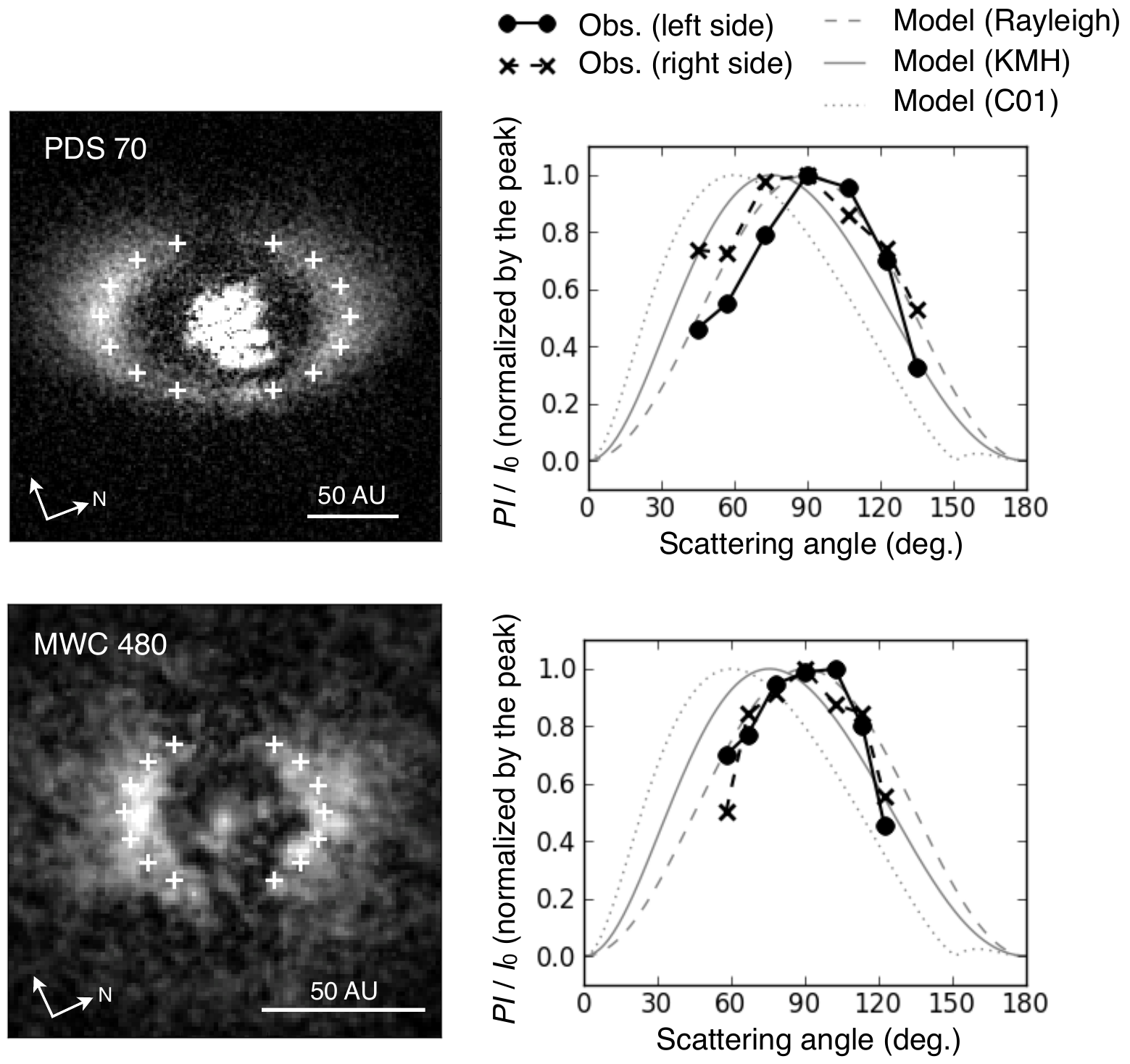}
\caption{($left$) Positions where we measure the $PI$ fluxes in PDS 70 and MWC 480. ($right$) The measured fluxes as a function of scattering angle. See Table \ref{tbl_objects} for the disk parameters required to derive the scattering angles. The figures also show the models for (1) the Rayleigh limit; (2) the interstellar dust model (KMH); and (2) the larger size distributions used by \citet{Cotera01} and \citet{Wood02b} to reproduce the scattered light observed in the HH 30 disk (C01). For each case the $PI$ flux is normalized by the peak of the curve.
\label{fig_phase_func}}
\end{figure*}

To investigate the validity of the dust model, we measure $PI$ fluxes at different positions in the PDS 70 and MWC 480 disks, and then plot them as a function of scattering angle. Figure \ref{fig_phase_func} shows the positions and fluxes compared with the following dust models: (1) the interstellar dust model described in Section 3 (KMH);
(2) the larger size distributions used by \citet{Cotera01} and \citet{Wood02b} to reproduce the scattered light observed in the HH 30 disk (C01); and (3) the Rayleigh limit. See \citet{Takami13} for detailed scattering properties for the KMH and C01 models. The plotted $PI$ flux is normalized by the peak flux for individual observations and models. The model profile peak at $90^\circ$, $75^\circ$, and $60^\circ$ for the Rayleigh limit, KMH and C01, respectively.

The measured $PI$ flux peaks at a scattering angle of $\sim 90^\circ$, and its distribution as a function of scattering angle is close to the Rayleigh limit. In this context, the dust grains in the scattering layer of these disks may be smaller than in the KMH model. Even so, the southwest side of the disk is brighter than the northeast side in PDS 70, which may be explained if the dust grains are larger than the Rayleigh limit.

The above analysis is accurate only if these disks have axisymmetric structures and even illumination. In practice, the structure or illumination of some disks may be non-axisymmetric (see Sections 4 and 5 for details). In contrast, the flux distribution for the two objects and two sides of the disk show a very similar trend in Figure \ref{fig_phase_func}, peaking at the scattering angle of $\sim 90 ^\circ$ and consistent with scattering on grains smaller than that in the KMH model. We thus conclude that non-axisymmetry in the disk structure or illumination does not significantly affect our discussion for grain sizes here.

To further investigate the dust properties one may alternatively measure the polarization, which is not affected by non-axisymmetric illumination \citep[e.g.,][]{Tanii12}. We note that, however, this approach faces several problems when one measures the polarization as $\sqrt{Q^2+U^2}/I$, where $I$, $Q$, $U$ are the Stokes parameters. First, it is significantly more difficult to accurately measure the $I$ flux distribution than the $PI$ flux distribution due to the bright stellar flux. Secondly, while the $PI$ flux is dominated by the singly scattered photons, photons with multiple scatterings contribute more significantly to the $I$ flux. As a result, the measured polarization would be 10--50 \% smaller than that predicted by single scattering on dust grains \citep{Takami13}. 




\clearpage



%
%
%

%
%

%

%
%

%
%
%

%
%
%

%
%
%

%
%
%

%

%
%
%

%
%
%

%
%
%

%
%
%








\clearpage

\begin{table}
\caption{Parameters \label{tbl_params}}
{\footnotesize
\begin{tabular}{ll}
\tableline\tableline
$A_{\rm pix}$	& Area corresponding to the HiCIAO pixel scale (9.5 mas) at the distance to the target\\
$A_{\rm Tel}$ 	& Area of the telescope mirror\\
$C$			& A constant to check the validity of one of the assumptions (see Equations 7 and 11 for definition) \\
$d$			& Distance to the disk from the observer \\
$dA$ 		& Differential area of the disk surface \\
$dA_{\rm out}$ & Differential area $dA$ projected onto the plane of the sky \\
$dA_*$ 		& Differential area $dA$ projected on the plane perpendicular to the light path from the star to the disk \\
$dr$			& Differential radius corresponding to $dA$ \\
$d \theta$ 	& Differential elevation angle corresponding to $dA$ \\
$d \phi$ 		& Differential azimuthal angle corresponding to $dA$ \\
$d \Omega_{\rm out}$ & Solid angle from the disk surface toward the observer\\
$h \nu $ 		& Photon energy\\
$I_{\nu; \rm out}\it (r,\phi)$ & Observed $PI$ intensity\\
$i$			& Disk inclination or the viewing angle \\
$L_\nu$ 		& Stellar luminosity per unit frequency \\
$n_{\nu;PI} (r,\phi)$	& Number of photons at a given pixel of the $PI$ intensity distribution \\
$n_{\nu;*}$	& Number of photons for the Stokes $I$ flux of the star \\
($PI/I_0$) 	& Fraction of the scattered $PI$ intensity relative to the incident $I_0$ flux (sr$^{-1}$) \\
$R (r,\phi)$ 	& Distance to the position of the scattering from the star\\
$r$			& Distance to the position of the scattering from the star projected onto the midplane (=$R \cos \theta$)\\
$S (r,\phi)$    	& $\equiv \sin \theta(r,\phi)$\\
$z$			& Height of the disk surface from the midplane\\
$\alpha$ 		& Angle of the differential area $dA$ with respect to the midplane \\
$\theta (r,\phi)$ 	& Elevation angle of the position of a scattering event from the midplane \\
$\phi$		& Azimuthal angle\\
$\Omega_{\rm Tel}$ & Solid angle corresponding to the telescope mirror \\
\tableline \tableline
\end{tabular}
}
\end{table}

\begin{table}
\caption{Model parameters for ID 3000949 of \citet{Robitaille07} \label{tbl_3000949}}
\begin{tabular}{lc}
\tableline\tableline
$\alpha$ 			& 2.138 \\
$\beta$          		& 1.138 \\
$h_{\rm 100 AU}$  	& 5.37 AU\\
$M_{\rm disk}$ 	& $1.06 \times 10^{-1}$ $M_\odot$	\\
Outer radius (AU) 	& 83.0 AU \\
Distance 			& 140 AU
\\ \tableline
\tableline
\end{tabular}
\end{table}

\begin{table}
\caption{$(PI/I_0)$ for different P.A.s \label{tbl_PI_I0}}
\begin{tabular}{cccc}
\tableline\tableline
Viewing angle $i$ & P.A. & typical scattering & $(PI/I_0)$ \\
($^\circ$) & ($^\circ$) & angle\tablenotemark{a} ($^\circ$) & str$^{-1}$ \\
\tableline
18.2	& 0		& 108.2	& $9.60 \times 10^{-3}$ \\
	& 45		& 102.9	& $1.04 \times 10^{-2}$ \\
	& 90		& 90.0	& $1.21 \times 10^{-2}$ \\
	& 135	& 77.1	& $1.28 \times 10^{-2}$ \\
	& 180	& 71.8	& $1.27 \times 10^{-2}$ \\
49.5	& 0		& 139.5	& $3.79 \times 10^{-3}$ \\
	& 45		& 129.6	& $5.61 \times 10^{-3}$ \\
	& 90		& 90.0	& $1.21 \times 10^{-2}$ \\
	& 135	& 50.4	& $1.03 \times 10^{-2}$ \\
	& 180	& 40.5	& $8.13 \times 10^{-3}$ \\
\tableline \tableline
\end{tabular}
\tablenotetext{a}{Calculated with the geometrically thin approximation}
\end{table}

\begin{table*}
\caption{$S_0$ for the best fit\tablenotemark{a} \label{tbl_S_0}}
\begin{tabular}{ccccccc}
\tableline\tableline
Model\tablenotemark{b} & Viewing angle $i$ & \multicolumn{5}{c}{P.A.} \\
           &($^\circ$)& $0^\circ$	& $45^\circ$	& $90^\circ$	& $135^\circ$	& $180^\circ$	\\
           \tableline
1	& 18.2	& $1.56 \times 10^{-1}$	& $1.58 \times 10^{-1}$	& $1.60 \times 10^{-1}$	& $1.64 \times 10^{-1}$	& $1.66 \times 10^{-1}$	\\
	& 49.5	& $1.57 \times 10^{-1}$	& $1.54 \times 10^{-1}$	& $1.64 \times 10^{-1}$	& $1.79 \times 10^{-1}$	& $1.83 \times 10^{-1}$	\\
2	& 18.2	& $1.96 \times 10^{-2}$	& $1.81 \times 10^{-2}$	& $1.75 \times 10^{-2}$	& $2.02 \times 10^{-2}$	& $2.19 \times 10^{-2}$	\\
	& 49.5	& $4.09 \times 10^{-2}$	& $3.08 \times 10^{-2}$	& $2.36 \times 10^{-2}$	& $4.07 \times 10^{-2}$	& $5.09 \times 10^{-2}$	\\
3	& 18.2	& $1.86 \times 10^{-2}$	& $1.72 \times 10^{-2}$	& $1.53 \times 10^{-2}$	& $1.84 \times 10^{-2}$	& $1.91 \times 10^{-2}$	\\
	& 49.5	& $3.73 \times 10^{-2}$	& $2.79 \times 10^{-2}$	& $2.20 \times 10^{-2}$	& $3.36 \times 10^{-2}$	& $4.25 \times 10^{-2}$	\\

\tableline
\tableline
\end{tabular}
\tablenotetext{a}{Defined at a projected radius of 5.31 AU from the star}
\tablenotetext{b}{1... the 300949 model of \citet{Robitaille07}; 2 ... Same as 1 but $\beta$=2, $\alpha$=3; 3 ... Same as 2 but with a disk gap (see text for details).}
\end{table*}

\begin{table*}
\caption{Maximum value of $C$\tablenotemark{a} \label{tbl_C_max}}
\begin{tabular}{ccccccc}
\tableline\tableline
Model\tablenotemark{b} & Viewing angle $i$ & \multicolumn{5}{c}{P.A.} \\
           &($^\circ$)& $0^\circ$	& $45^\circ$	& $90^\circ$	& $135^\circ$	& $180^\circ$	\\
           \tableline
1	& 18.2	& 0.1	& 0.1	& 0.09	& 0.08	& 0.08	\\
	& 49.5	& ---		& ---		& 0.1	& ---		& ---		\\
2	& 18.2	& 0.6	& 0.5	& 0.5	& 0.4	& 0.4	\\
	& 49.5	& ---	& ---	& 0.7	& ---	& ---	\\
3	& 18.2	& 0.8	& 0.8	& 0.8	& 0.8	& 0.7	\\
	& 49.5	& ---	& ---	& 1.1	& ---	& ---	\\

\tableline
\tableline
\end{tabular}
\tablenotetext{a}{See Equations (7) and (11) for definition. We tabulate values for Models 2 and 3 and $i$=49.5$^\circ$ only for P.A.=90$^\circ$ (see text).}
\tablenotetext{b}{1... the 300949 model of \citet{Robitaille07}; 2 ... Same as 1 but $\beta$=2, $\alpha$=3; 3 ... Same as 2 but with a disk gap (see text for details).}
\end{table*}


\begin{table*}
\begin{center}
\caption{Objects\tablenotemark{a}\label{tbl_objects}}
{\scriptsize
\hspace*{-1cm}
\begin{tabular}{cccccccc}
\tableline\tableline
Object 		& Distance & Stellar mass\tablenotemark{b} & Structures & Inclination  & P.A. of major axis & Peak $PI/I_*$ & Reference \\
           		& (pc)	& ($M_\odot$) & observed  & ($^\circ$)   & ($^\circ$) & (pixel$^{-1}$) \\
\tableline
SAO	206462	& 140	& 1.7 & spirals		& 11	& 55	&$\sim 1 \times 10^{-6} $ 	& \citet{Muto12} \\
MWC 758		& 200 	& 2.2& spirals		& 21\tablenotemark{c}	& 65		&	$\sim 9 \times 10^{-7} $ & \citet{Grady13} \\
2MASS J16042165--2130284		& 145 &1.0& ring		& 10	& 76	&	$\sim 2 \times 10^{-6} $& \citet{Mayama12} \\
PDS 70		& 140 &	0.8& ring		& 50	& 159	&	$\sim 6 \times 10^{-7} $& \citet{Hashimoto12} \\
MWC 480		& 140 &	2.3& uniform	& 38\tablenotemark{c} & 155\tablenotemark{d} &	$\sim 1 \times 10^{-7} $ & \citet{Kusakabe12} \\
\tableline
\tableline
\end{tabular}
}
\tablenotetext{a}{Uncertainties for distance, inclination and P.A. are omitted in this table as these are not described in the references in a consistent manner.}
\tablenotetext{b}{References --- \citet{Muller11} for SAO 206462; \citet{Mannings97} for MWC 758 and 480; \citet{Preibisch99} for 2MASS J1604; and \citet{Riaud06} for PDS 70.}
\tablenotetext{c}{Based on millimeter interferometry by \citet{Isella10b} for MWC 758; \citet{Pietu07} for MWC 480.}
\tablenotetext{d}{Measured this time base on the symmetry of the observed $PI$ distribution shown in Appendix A.}
\end{center}
\end{table*}


\begin{table*}
\begin{center}
\caption{Parameters for the revised surface geometry\label{tbl_params_for_revised_z}}
\begin{tabular}{ccccc}
\tableline\tableline
Object 		& $r_1$ & $z_1$ & $S_0$ & $f$ \\
           		& (AU)	& (AU)  & (Case A) & (Case B)\\
\tableline
SAO	206462	& 110	&10.3	&0.000--0.059		&0.36--1.00\\
MWC 758		& 120	&5.3		&0.000--0.017		&0.56--1.00\\
2MASS J16042165--2130284		& 120 & 20.8	&0.001--0.074	&0.57--1.00\\
PDS 70		& 120	&7.0		&0.001, 0.019		&0.99, 0.68\\
MWC 480		& 120	&0.63	&$6.1 \times 10^{-4}$, $0.7 \times 10^{-4}$	&0.88, 0.99\\
\tableline
\tableline
\end{tabular}
\end{center}
\end{table*}

\end{document}